\documentclass[aps,prc,letterpaper,showpacs,12pt,floatfix,preprint]{revtex4-1}
\pdfoutput=1
\usepackage{natbib}
\usepackage{amsmath}
\usepackage{graphicx}
\usepackage{hyperref}
\usepackage{multirow}
\usepackage{color}
\usepackage[utf8]{inputenc}
\usepackage{bm}

\begin{document}

\preprint{\vbox{\hbox{JLAB-THY-20-3132}}}

\title{Neutral-Current Neutrino Scattering from the Deuteron}

\author{Sabine Jeschonnek$^{(1)}$, J. W. Van Orden$^{(2,3)}$, and T. W. Donnelly$^{(4)}$}

\affiliation{\small \sl 
	(1) The Ohio State University, Physics
Department, Lima, OH 45804\\
(2) Department of Physics, Old Dominion University, Norfolk, VA
23529\\ (3) Jefferson Lab, \footnote{Notice: Authored by Jefferson Science Associates, LLC under U.S. DOE Contract No. DE-AC05-06OR23177. The U.S. Government retains a non-exclusive, paid-up, irrevocable, world-wide license to publish or reproduce this manuscript for U.S. Government purposes.} 
\\ 12000 Jefferson Avenue, Newport
News, VA 23606\\
(4) Massachusetts Institute of Technology, Cambridge, MA 02139 
 }

\date{\today}

\begin{abstract}

Neutral-current neutrino scattering from the deuteron leading to proton-neutron final states is considered using an approach that incorporates relativistic dynamics and consequently provides robust modeling at relatively high energies and momenta. In this work the focus is placed on the fully exclusive reaction where both the proton and neutron in the final state are assumed to be detected. Accordingly, the incident neutrino energy, the neutrino scattering angle and the scattered neutrino’s energy can all be reconstructed. It is shown that for specific choices of kinematics the reaction proceeds mainly via scattering from the proton, while for other choices of kinematics it proceeds mainly from the neutron. Specific asymmetries are introduced to focus on these attributes. Measurements in both regions have the potential to yield valuable information on the nucleon’s electroweak form factors at momentum transfers up to a (GeV/c)$^2$. In particular, the cross sections are shown to be very sensitive to the isoscalar axial-vector form factor, and sensitive but less so to the magnetic strangeness form factor. Comparisons with other reactions, specifically charge-changing neutrino reactions and both parity-conserving and -violating electron scattering, have the potential to provide new ways to test the Standard Model.

\end{abstract}

\maketitle


\section{Introduction}\label{sec:intro}

The present study involves an investigation of neutral-current neutrino and anti-neutrino scattering from the deuteron, especially at high energies of order several GeV, namely those of present and anticipated neutrino facilities. One motivation for such a study is to explore the possibilities for learning more about the form factors of the nucleon, $N=p$ or $n$, over a wide range of momentum transfers. Of course, the electromagnetic form factors of the proton may be accessed through parity-conserving electron scattering and those of the neutron typically involve elastic or inelastic electron scattering from the deuteron. However, the other form factors that enter when the weak interaction is involved are harder to determine. One way this is done is via parity-violating electron scattering from the proton where constraints have been placed on the proton's axial-vector and strangeness form factors, although the present constraints, while impressive, are not definitive. Other parity-violating electron scattering reactions also play a role, specifically coherent elastic parity-violating electron scattering --- for instance, from $^4$He or $^{208}$Pb --- where the electric strangeness form factor of the nucleon enters in the PV asymmetry, albeit at rather low values of the momentum transfer (for discussions on the last point the reader is directed to \cite{Moreno:2015bta}).

Neutrino reactions with the proton or deuteron also depend on a more complete set of nucleon form factors and accordingly, these too play a role. For charge-changing neutrino reactions (CC$\nu$) having an incident beam of neutrinos of given flavor, $\nu_\ell$ with $\ell = e$, $\mu$ or $\tau$, going to a negatively charged lepton of the same flavor, $\ell^-$, scattering from the proton leading to a neutron in the final state cannot occur --- it does, of course, for anti-neutrinos with positively charged leptons $\ell^+$ in the final state --- whereas it can occur for CC$\nu$ reactions on neutrons leading to protons in the final state. That is, the CC$\nu$ reactions that can occur below pion production threshold are ${\bar\nu}_\ell + p \rightarrow \ell^+ + n$ and $\nu_\ell + n \rightarrow \ell^- + p$. Unfortunately, practical free neutron targets do not exist and thus, when neutrinos constitute the beam and these elementary reactions provide the focus, one must employ nuclei which contain neutrons. A special case of this type is that of the deuteron, since, of all nuclei, one has the most robust ability to model it, not just at low energies, but at relatively high energies where relativistic approaches must be pursued; this aspect is discussed in more detail later in the paper and draws on studies performed previously. Of the latter type we mention the relatively recent work on CC$\nu$ disintegration of the deuteron going into $pp$ in the final state \cite{VanOrden:2017uyy} (see also \cite{Moreno:2014kia,Moreno:2015nsa}).

Finally, in addition to CC$\nu$ reactions where only isovector form factors enter, one can also have neutral-current neutrino or anti-neutrino scattering reactions (NC$\nu$ generically), which include elastic scattering of neutrinos or anti-neutrinos from protons or neutrons, namely, $\nu_\ell + p \rightarrow \nu_\ell + p$, ${\bar\nu}_\ell + p \rightarrow {\bar\nu}_\ell + p$ and the analogous reactions on neutrons. Again, in the absence of practical free neutron targets, the deuteron plays a special role. Specifically, the present study is focused on the NC$\nu$ disintegration reactions $\nu_\ell + ^2$H$\rightarrow \nu_\ell + p + n$ and ${\bar\nu}_\ell + ^2$H$\rightarrow {\bar\nu}_\ell + p + n$ where {\em both} the proton and neutron are assumed to be detected. 

Thus two classes of NC$\nu$ reactions have the potential to provide new information on the form factors of the proton and neutron, namely, elastic neutrino and anti-neutrino scattering from the proton and the above neutrino-disintegration reactions on deuterium. 

\begin{figure}
	\centering
	\includegraphics[width=8cm]{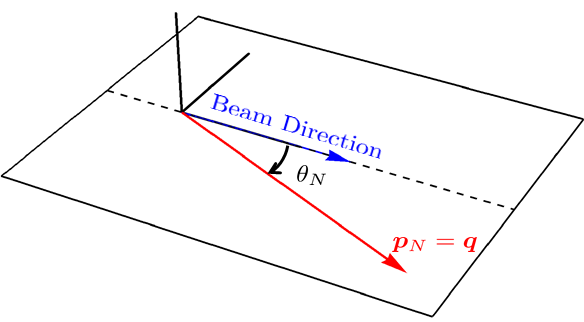} 	\includegraphics[width=8cm]{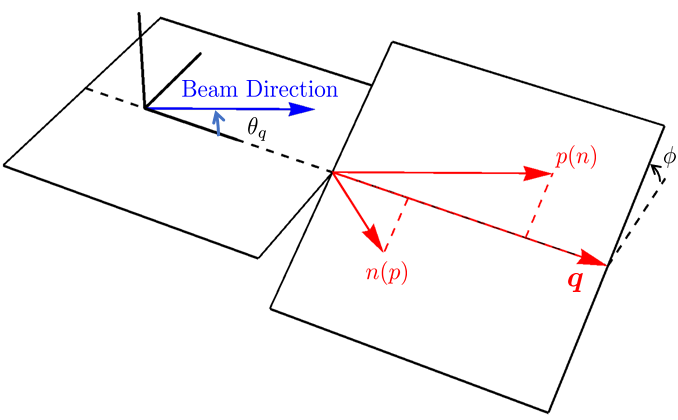} 			
	\caption{Schematic figures showing the reactions of interest in the present work. On the left-hand side one has elastic scattering from a nucleon which, after the scattering of the neutrino or anti-neutrino, recoils with momentum $\boldsymbol{p}_N$, namely, along the momentum transfer $\boldsymbol{q}$. On the right-hand side one has the NC disintegration of deuterium into a final-state $pn$ pair. The total momentum of this pair is again given by the momentum transfer $\boldsymbol{q}$, as shown.}
	\label{fig:n_and_d_coordinates}
\end{figure}
In Fig.~\ref{fig:n_and_d_coordinates} we show schematic representations of these classes of reactions. For elastic scattering from the nucleon (left-hand panel of the figure) the nucleon recoils in the direction of the momentum transferred from the neutrino to the nucleon. Similarly, for the disintegration of the deuteron into proton plus neutron (the right-hand panel in the figure) momentum $\boldsymbol{q}$ and energy $\omega$ are transferred to the $pn$ system. By detecting {\em both} nucleons one can reconstruct the magnitude of the three-momentum transfer $q=|\boldsymbol{q}|$, its direction with respect to the incident neutrino beam, characterized by an angle $\theta_q$, and the energy transfer $\omega$. As discussed in detail later in the paper, this then allows one to reconstruct the entire lepton-scattering kinematics, namely, the neutrino beam energy, the neutrino scattering angle and the scattered neutrino energy. Thus one has a very special set of circumstances: by selecting events having no final-state charged lepton and no pions one can, in effect, have a very selective, well-modeled neutrino-disintegration reaction to study. And, since the momentum transfers can be quite large, of order a (GeV/c)$^2$ as will be shown later, this opens new possibilities for exploring the nucleon's electroweak form factors.

As discussed later in the paper, for the NC$\nu$ disintegration reaction there are kinematic regions where scattering from the proton in the two-body systems is dominant, hence yielding information on the proton form factors which can be cross-checked with results for elastic scattering for the proton, but also (importantly) regions where scattering from the neutron is dominant and information on the neutron form factors becomes accessible. Since, the proton and neutron form factors may be rewritten in terms of isoscalar and isovector nucleon form factors, this has the potential to open new possibilities to explore the entire set of electroweak form factors of the nucleon.

The paper is organized in the following way: in Sect.~\ref{sec:nucleon} the vector and axial-vector form factors for the proton and neutron in the Standard Model are summarized along with parametrizations for these functions that have been employed in past work \cite{Donnelly:1978tz,Musolf:1992xm,Musolf:1993tb}, together with expressions for the elastic scattering NC$\nu$ cross sections. There we also introduce specific $pn$-asymmetries for both neutrinos and anti-neutrinos that prove useful in the discussions to follow. In Sect.~\ref{sec:deuteron} the NC$\nu$ disintegration reactions involving the deuteron introduced above are explored, building on previous work. Here both the basic formalism for exclusive reactions (detection of both the proton and neutron in the final state), as well as typical results for the cross sections and asymmetries (to be defined there) are discussed. Finally in Sect.~\ref{sec:summary} we present a summary of what has been learned from these investigations.

For a recent overview of present and future experimental potential in the area of neutrino reactions with nucleons and nuclei the reader is directed to \cite{Alvarez-Ruso:2017oui}.

\section{Neutrino Scattering from the Nucleon}\label{sec:nucleon}

\begin{figure}[h]
	\centering
	\includegraphics[height=2in]{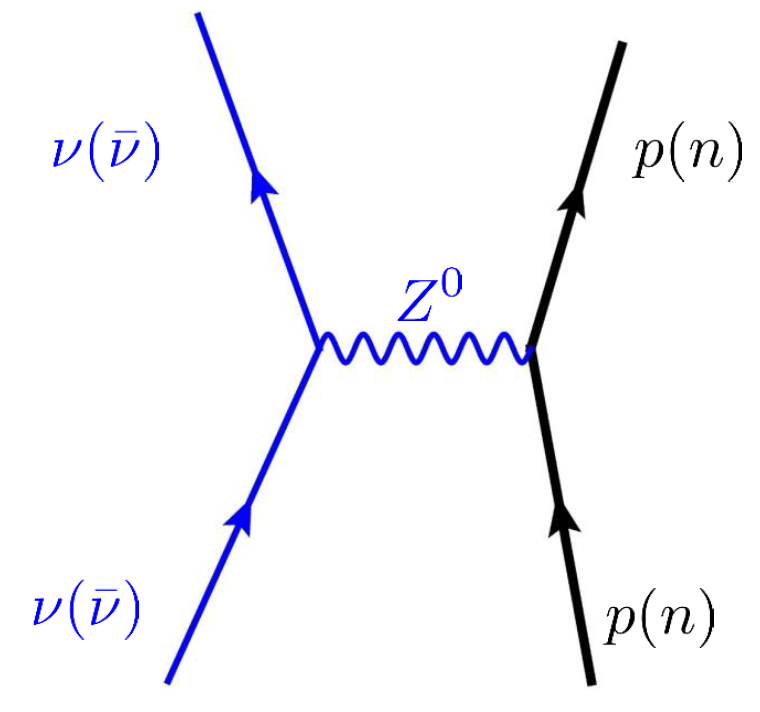} 
	\caption{Diagram representing NC$\nu$ scattering from a nucleon.}
	\label{fig:nucleon_diagram}
\end{figure}

Neutral-current neutrino or anti-neutrino scattering from a nucleon is represented by Fig.~\ref{fig:nucleon_diagram}.
Later in this section we present explicit forms for the scattering cross sections; however, we begin by summarizing the complete set of single-nucleon electroweak form factors. Here we draw upon old work \cite{Donnelly:1978tz,Musolf:1992xm,Musolf:1993tb} including the parametrizations used previously.

\subsection{Electroweak Form Factors of the Nucleon}\label{subsec:formN}

The effective coupling of the $Z^0$ to nucleons (the right-hand part of the figure) involves the electroweak current discussed in the above-cited work which in turn may be written in a standard form involving vector and axial-vector currents. The vector currents for the proton may be written in terms of Sachs form factors as
\begin{align}
	\widetilde{G}_{E,M}^{p}(\tau) =&(1-2\sin^2\theta_W)G_{E,M}^{(1)}(\tau)-2
	\sin ^{2}\theta _{W}G_{E,M}^{(0)}(\tau)-G_{E,M}^{(s)}(\tau)\nonumber\\
	=&\frac{1}{2}\left((1-4\sin ^{2}\theta _{W})
	G_{E,M}^p(\tau)-G_{E,M}^n(\tau)-2G_{E,M}^{(s)}(\tau)\right) 
\end{align}
and for the neutron as
\begin{align}
	\widetilde{G}_{E,M}^{n}(\tau)=&
	-(1-2\sin^2\theta_W)G_{E,M}^{(1)}(\tau)
	-2\sin^2\theta_ W G_{E,M}^{(0)}(\tau)
	-G_{E,M}^{(s)}(\tau)\nonumber\\  
	=&\frac{1}{2}
	\left((1-4\sin ^{2}\theta _{W})G_{E,M}^n(\tau)
	-G_{E,M}^p(\tau)-2G_{E,M}^{(s)}(\tau)\right)\,,
\end{align}
where $\tau\equiv |Q^2|/4m_N^2$ with $Q^2=\omega^2 - q^2 <0$, and we have used 
\begin{equation}
G^{(0)}_{E,M}(\tau) = G^p_{E,M}(\tau) + G^n_{E,M}(\tau)
\end{equation}
and
\begin{equation}
G^{(1)}_{E,M}(\tau) = G^p_{E,M}(\tau) - G^n_{E,M}(\tau)  
\end{equation}
for the isoscalar and isovector electromagnetic form factors, namely those with $T=0$ or $1$, respectively. These expressions contain the usual Standard Model mixing of the isoscalar and isovector electromagnetic form factors via the expressions containing the term $\sin^2\theta_W$. In addition,  possible isoscalar strangeness form factors $G^{(s)}_{E,M}(\tau)$ are included.

The nucleon axial-vector form factors for the proton and neutron are given by
\begin{equation}
\widetilde{G}_{A}^{p}(\tau) =-G_{A}^{(1)}(\tau
)+G_{A}^{(s)}(\tau) 
\end{equation}
and
\begin{equation}
\widetilde{G}_{A}^{n}(\tau) =G_{A}^{(1)}(\tau)
+G_{A}^{(s)}(\tau)
\,, 
\end{equation}
where $G^{(1)}_A(\tau)$ is the usual isovector axial-vector form factor that appears in charged-current reactions and $G^{(s)}_A(\tau)$ is an isoscalar axial-vector strangeness form factor \cite{Musolf:1993tb}.

For the results presented in this paper the electromagnetic currents are from a fit to electron scattering data using a vector dominance model \cite{Lomon:2002jx,Lomon:2006xb,Crawford:2010gv}.  Following \cite{Musolf:1992xm,Musolf:1993tb} the axial-vector and strangeness form factors are chosen to involve simple dipole forms 
\begin{align}
G_A^{(1)}(\tau)=&g_A G_D(\tau)\nonumber\\
G_E^{(s)} (\tau) =&  \rho_s \tau G^D_V(\tau)\nonumber\\
G_M^{(s)} (\tau) =&  \mu_s G^D_V(\tau)\nonumber\\
G_A^{(s)} (\tau) =&  g^s_{A} \tau G^D_A(\tau)\,,
\end{align}
where the dipole form factors are $G^D_V(\tau) = ( 1 + 4.97 \tau)^{-2}$ and $G^D_A(\tau) = ( 1 + 3.32 \tau)^{-2}$.
In this form the strangeness content of the nucleons is characterized by the coupling constants $\rho_s$, $\mu_s$ and $g^s_A$.

The analysis of parity-violating electron scattering experiments has produced constraints on the possible values for these couplings \cite{Gonzalez-Jimenez:2014bia,GonzalezJimenez:2011fq,Moreno:2014ksa} which we employ in the present work to provide a ``reasonable'' representation for the lesser-known electroweak form factors. Additionally, lattice QCD (LQCD) calculations of strangeness contributions to the electromagnetic form factors \cite{Sufian:2016vso,Djukanovic:2019jtp} have been performed and a calculation of the strangeness contribution to the axial form factor using input from LQCD and experiment has recently been published \cite{Sufian:2018qtw}. A complete {\it ab initio} LQCD calculation of the weak axial vector form factor is still needed however \cite{Kronfeld:2019nfb}.

The differences seen in the isospin dependences of the form factors for protons and neutrons suggests the possibility that isospin asymmetries constructed from neutral-current neutrino interactions with nucleons could show increased sensitivity to these couplings and thereby yield further constraints on these couplings. The obvious problem with this, as discussed in the Introduction, is that no practical free neutron targets exist, and usually the neutron properties are obtained from reactions with the deuteron. In this paper we will demonstrate that neutral-current neutrino scattering from deuterium can, with a careful choice of kinematics, be used to produce the nucleon isospin asymmetries. Calculations demonstrating this approach will use relativistic deuteron models that were developed to describe the $^2$H$ (e,e'p)$ reaction \cite{Jeschonnek:2008zg,Jeschonnek:2009tq,Jeschonnek:2009ds,Ford:2013uza} and the charge-changing neutrino reaction with the deuteron  \cite{Moreno:2014kia}.

Obviously, our main goal is to look at neutrino scattering from the deuteron, as this will give us information on both the proton and the neutron strangeness content. We shall, however, start out with an exploration of the sensitivity of the neutrino-nucleon cross section to the strangeness parameters in order to get our kinematical bearings, 
pretending for the moment that we have free neutron targets available. We proceed in the next sub-section with a summary of the kinematics involved.

\subsection{NC$\nu$ Elastic Scattering Kinematics}\label{subsec:kineN}

We start out by establishing our notation and recalling a few kinematical relations. In the case of NC$\nu$ scattering, both incoming and outgoing leptons are neutrinos of the same flavor with (as assumed in this work) mass $m=0$. First, 
\begin{equation}
K^\mu=(k,\bm{k})
\end{equation}
is the initial neutrino four-momentum, and the scattered neutrino four-momentum is given by
\begin{equation}
{K'}^\mu=(k',\bm{k}') \,.
\end{equation}
Then the four-momentum transfer is
\begin{equation}
Q^\mu=K^\mu-{K'}^\mu=(k-k',\bm{k}-\bm{k}')=(\omega,\bm{q})
\end{equation}
and is spacelike, $Q^2 = \omega^2 - q^2 < 0$.
The four-momentum of the target nucleon can be written in its rest frame as
\begin{equation}
P^\mu=(m_N,\bm{0})\,,
\end{equation}
while the four-momentum of the detected nucleon is
\begin{equation}
{P'}^\mu=(\sqrt{p_N^2+m_N^2},\bm{p}_N)\,.
\end{equation}

Energy conservation requires that
\begin{equation}
m_N+k-k'-\sqrt{p_N^2+m_N^2}=0
\end{equation}
or
\begin{equation}
k-k'-T_N=0\,,  \label{eq:EC_single}
\end{equation}
where the relativistic kinetic energy of the nucleon is defined as
\begin{equation}
T_N=\sqrt{p_N^2+m_N^2}-m_N\,.
\end{equation}
This implies that the energy transfer is given by
\begin{equation}
\omega=T_N\,.
\end{equation}
Furthermore, momentum conservation requires that
\begin{equation}
\bm{k}-\bm{k'}-\bm{p}_N=0\label{eq:MC_single}
\end{equation}
or
\begin{equation}
\bm{p}_N=\bm{q}\,.
\end{equation}
From (\ref{eq:MC_single}) one has that
\begin{equation}
\bm{k}'=\bm{k}-\bm{p_N}\,,
\end{equation}
and, using this in (\ref{eq:EC_single}), we find that
\begin{equation}
k-\sqrt{k^2-2kp_N\cos\theta_N+p_N^2}-T_N=0\,,
\end{equation}
where $\theta_N$ is the angle between $\bm{k}$ and $\bm{p}_N$. Solving this for $k$ gives
\begin{equation}
k=\frac{m_N T_N}{p_N\cos\theta_N-T_N }\,,\label{eq:k_single}
\end{equation}
and the energy of the scattered neutrino is then
\begin{equation}
k'=-\frac{T_N(m_N + T_N - p_N\cos\theta_N)}{p_N\cos\theta_N-T_N}\,.\label{eq:kp_single}
\end{equation}
Note that (\ref{eq:k_single}) and (\ref{eq:kp_single}) become singular when $T_N=p_N\cos\theta_N$.

Since $x=1$ for elastic scattering, the square of the four-momentum transfer is given by
\begin{equation}
|Q^2|=2 m_N\omega=2 m_N T_N\,.
\end{equation}
The square of the three-momentum transfer is given by
\begin{equation}
q^2=k^2-2 k k'\cos\theta_l+ {k'}^2=p_N^2\,.
\end{equation}
Using (\ref{eq:k_single}) and (\ref{eq:kp_single}), the neutrino scattering angle can be obtained from
\begin{equation}
\cos\theta_l=\frac{m_N T_N + p_N T_N\cos\theta_N - p_N^2\cos^2\theta_N}
{T_N(m_N+T_N-p_N\cos\theta_N)}\,.
\end{equation}
Finally, it is convenient to define a dimensionless variable for the square of the four-momentum transfer as
\begin{equation}
\tau=\frac{|Q^2|}{4m_N^2}=\frac{T_N}{2m_N}\,.
\end{equation}
This completes the basic kinematical relationships needed when treating elastic neutrino scattering from the nucleon. Next we turn to a summary of the essentials for the single-nucleon cross section.

\subsection{Single-Nucleon Cross Section}\label{subsec:singleN}

The double-differential cross section for neutral-current neutrino scattering from a nucleon is given by (for general discussions of semi-inclusive neutrino reactions with nuclei see \cite{Moreno:2014kia})
\begin{align}
\frac{d\sigma}{dp_Nd\Omega_N}=&            
\frac{G^2 \cos^2\theta_C m_N p_N^2v_0}
{16\pi^2  \sqrt{p_N^2+m_N^2}(p_N\cos\theta_N-T_N)}\frac{P(k)}{k}
\left[\hat{V}_{CC}(w_{CC}^{VV(I)}+w_{CC}^{AA(I)})\right.\nonumber\\
&+2\hat{V}_{CL}(w_{CL}^{VV(I)}+w_{CL}^{AA(I)})
+\hat{V}_{LL}(w_{LL}^{VV(I)}+w_{LL}^{AA(I)})\nonumber\\
&+\left.\hat{V}_T(w_{T}^{VV(I)}+w_{T}^{AA(I)})+\chi\hat{V}_{T'}w^{VA(I)}_{T'}\right]\,,
\end{align}
where P(k) is a flux weighting factor that has the shape of an experimental neutrino flux distribution arbitrarily normalized to one, the nucleon response functions are
\begin{align}
w_{CC}^{VV(I)}=&(1+\tau)\widetilde G_E^2(\tau) \nonumber\\
w_{CL}^{VV(I)}=& \sqrt{\tau(1+\tau)}\widetilde G_E^2(\tau)\nonumber\\
w_{LL}^{VV(I)}=& \tau\widetilde G_E^2(\tau)\nonumber\\
w_{T}^{VV(I)}=&2\tau \widetilde G_M^2(\tau) \nonumber\\
w_{CC}^{AA(I)}=&\tau\widetilde G_A^2(\tau) \nonumber\\
w_{CL}^{AA(I)}=&\sqrt{\tau  (1+\tau)}\widetilde G_A^2(\tau) \nonumber\\
w_{LL}^{AA(I)}=&(1+\tau)\widetilde G_A^2(\tau) \nonumber\\
w_{T}^{AA(I)}=&2(1+\tau) \widetilde G_A^2(\tau) \nonumber\\
w^{VA(I)}_{T'}=&4\sqrt{\tau 
	(1+\tau)} \widetilde G_A(\tau) \widetilde G_M(\tau)\,. 
\end{align}
and the kinematic factors are 
\begin{align}
	&v_0= 4|\bm{k}||\bm{k}'|\cos^2	\frac{\theta_l}{2} \nonumber\\
	&\hat{V}_{CC}= 1 \nonumber\\
	&\hat{V}_{CL}= -\frac{\omega}{|\bm{q}|}\nonumber\\
	&\hat{V}_{LL}=\frac{\omega^2}{\bm{q}^2}\nonumber\\
	&\hat{V}_{T}=\left( \frac{|Q^2|}{2\bm{q}^2}+\tan^2\frac{\theta_l}{2}\right)\nonumber\\
	&\hat{V}_{TT}= -\frac{|Q^2|}{2\bm{q}^2} \nonumber\\
	&\hat{V}_{TC}= -\frac{1}{\sqrt{2}}\sqrt{\frac{Q^2}{\bm{q}^2}+\tan^2\frac{\theta_l}{2}}\nonumber\\
	&\hat{V}_{TL}= -\frac{1}{\sqrt{2}}\frac{\omega}{|\bm{q}|}\sqrt{\frac{|Q^2|}{\bm{q}^2}+\tan^2\frac{\theta_l}{2}}\nonumber\\
	&\hat{V}_{T'}= \tan\frac{\theta_l}{2}\sqrt{\frac{|Q^2|}{\bm{q}^2}+\tan^2\frac{\theta_l}{2}}\nonumber\\
	&\hat{V}_{TC'}= -\frac{1}{\sqrt{2}}\tan\frac{\theta_l}{2}\nonumber\\
	&\hat{V}_{TL'}= \frac{1}{\sqrt{2}}\frac{\omega}{|\bm{q}|}\tan\frac{\theta_l}{2}\,.\label{eq:kinematic_factors}
\end{align}

It is convenient to rewrite the cross section as
\begin{align}
\left(\frac{d\sigma}{dp_Nd\cos\theta_N}\right)_\chi=&\frac{G^2\cos^2\theta_c}{8\pi }\frac{P(k)}{k}\left(\Upsilon_\mathrm{CC+CL+LL}+\Upsilon_\mathrm{T}+\chi\Upsilon_\mathrm{T'}\right) \,,  \label{eq:nucleon_cross_section}         
\end{align}
where
\begin{align}
\Upsilon_\mathrm{CC+CL+LL}=&\frac{8  m_N^3
	p_N^2 \sin^2\theta_N
	T_N\tilde{G}_E^2(\tau)}{(m_N+T_N) (2
	m_N+T_N)
	(p_N\cos\theta_N
	-T_N)^3}\label{eq:Upsilon_1}
\end{align}
\begin{align}
\Upsilon_\mathrm{T}=&\frac{4 m_N p_N^2
	T_N \left((m_N+T_N)\cos^2\theta_N
	-2
	p_N\cos\theta_N
	+m_N+T_N\right)
}{(m_N+T_N) (2 m_N+T_N)
	(p_N\cos\theta_N -T_N)^3}\nonumber\\
&\times	\left(T_N\tilde{G}_M^2(\tau)
+ (2
m_N+T_N)\tilde{G}_A^2(\tau)\right)\label{eq:Upsilon_2}
\end{align}
\begin{align}
\Upsilon_\mathrm{T'}&=\frac{8  m_N
	T_N^2 (2
	m_N-p_N\cos\theta_N+T_N
	)\tilde{G}_A(\tau) \tilde{G}_M(\tau)}{(m_N+T_N) (p_N\cos\theta_N
	-T_N)^2}\label{eq:Upsilon_3}
\end{align}
with
$\chi=-1$ for neutrino scattering and $\chi=1$ for anti-neutrino scattering.

Specifically, for backward scattering, where $\theta_l=\pi$, $\bm{q}=\bm{p}_N$ is parallel to the beam direction. So, $\theta_N=0$, and in this limit
\begin{align}
&\lim_{\theta_N\rightarrow 0}\Upsilon_{CC+CL+LL}=0\nonumber\\
&\lim_{\theta_N\rightarrow 0}\Upsilon_T=\frac{8 m_N T_N
	(m_N-p_N+T_N)
	\left(T_N\tilde{G}_M^2(\tau)
	+ (2
	m_N+T_N)\tilde{G}_A^2(\tau)\right)}{(m_N+T_N) (2 m_N+T_N)
	(p_N
	-T_N)^3}\nonumber\\	
&\lim_{\theta_N\rightarrow 0}\Upsilon_{T'}=\frac{8  m_N
	T_N^2 (2
	m_N-p_N+T_N)\tilde{G}_A(\tau) \tilde{G}_M(\tau)}{(m_N+T_N)
	(p_N-T_N)^2}\label{eq:nucleon_0} \,.
\end{align}
The first of these indicates that the contribution from the $CC$, $CL$ and $LL$ responses vanish for backward scattering. This means that there is no contribution to the cross section from $\widetilde G_E$. The second is the result of contributions from the vector-vector and axial-axial contributions to the transverse response. These naturally contain contributions proportional to $\widetilde G_M^2$ and $\widetilde G_A^2$. The third term is the contribution from the vector-axial interference contribution to the $T'$ response and is proportional $\widetilde G_M\widetilde G_A$. Since $\chi=-1$ for neutrino scattering and $\chi=1$ for anti-neutrino scattering, the second and third contributions to the cross section can be separated by measuring cross sections for both neutrino and anti-neutrino scattering. Furthermore, comparisons of neutrino (anti-neutrino) scattering from both protons and neutrons may allow for increased sensitivity to strangeness form factors due to the isospin dependence of the form factors.

In particular, the isospin dependence of the nucleon form factors suggests that the sensitivity to the parameters $\rho_s$, $\mu_s$ and $g_A^s$ might be increased by forming a ratio of isovector to isoscalar combinations of the nucleon cross sections defined as
\begin{equation}
\mathcal{A}_\chi=\frac{ \left(d\sigma/dp_Nd\cos\theta_N\right)_\chi^p- 
	\left(d\sigma/dp_Nd\cos\theta_N\right)_\chi^n} {\left(d\sigma/dp_Nd\cos\theta_N\right)_\chi^p+ \left(d\sigma/dp_Nd\cos\theta_N\right)_\chi^n}\,. \label{eq:isospin_asymmetry}
\end{equation}
This ratio should produce cancellation of some systematic errors; for example, the neutrino flux factors cancel in this ratio. Note that this asymmetry is defined for both neutrinos and the corresponding anti-particles. In order to conserve space in this paper we present only cross sections and asymmetries for neutrinos although we have calculated the corresponding quantities for the anti-particles. It is clear that other asymmetries involving combinations of cross sections for both isospins and both values of $\chi$ can also be defined. 

In order to get an idea of the size of effects and the sensitivity to the strangeness parameters that we can expect without any nuclear effects, we first
present numerical results for single-nucleon targets. As a theoretical demonstration of this approach we will examine the sensitivity of the isospin asymmetry to variations in $\rho_s$, $\mu_s$ and $g_A^s$ in the next section before proceeding to consideration of whether or not a similar isospin asymmetry can be produced from CC$\nu$ neutrino scattering from the deuterium cross sections alone.

\subsection{Single-Nucleon Sensitivity to the Strangeness Parameters}\label{subsec:strangeN}

We will now consider the sensitivity of the nucleon cross sections to variations of values of the parameters $\rho_s$, $\mu_s$ and $g_A^s$. For this purpose we have chosen five values of each of the parameters over ranges that contain the probability ellipses obtained from fits to parity-violating electron scattering data in \cite{Gonzalez-Jimenez:2014bia} and the results of recent lattice QCD calculations \cite{Sufian:2018qtw,Djukanovic:2019jtp}. The chosen parameters are listed in Table \ref{tab:parameters}.
\begin{table}
	\caption{Strangeness parameters used in testing the variation of cross sections and asymmetries.}
	\begin{tabular}{p{1.5cm} p{1.5cm} p{.75cm} }\hline\hline
		$\rho_s$ & $\mu_s$ & $g^A_s$ \\ \hline
		-0.3 & -0.1 & -0.4 \\
		0.0 & 0.0 & -0.3 \\
		0.3 & 0.1 & -0.2 \\
		0.6 & 0.2 & -0.1 \\
		0.9 & 0.3 & 0.0\\ \hline\hline	
	\end{tabular}\label{tab:parameters}
\end{table}
In order to limit the number of figures showing the results of the variation in parameters, we produce figures where two of the parameters are fixed at the central values given in the third line of Table \ref{tab:parameters} and the third is varied. Each parameter is varied over equal steps to allow easy interpolation to intermediate values using the resulting figures. All cross sections are calculated using a flux factor having the shape of the DUNE flux arbitrarily normalized to 1.

\begin{figure}
	\centering
	\includegraphics[width=8cm]{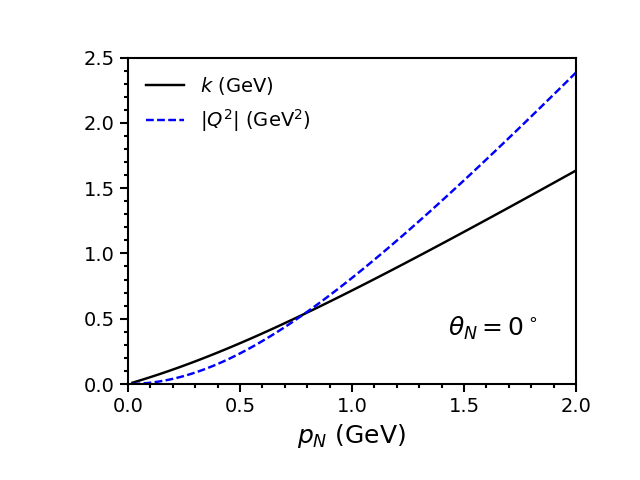} 	
	\caption{Values of $k$ and $|Q^2|$ for neutral-current neutrino scattering from nucleons at $\theta_N=0^\circ$ as a function of the nucleon momentum $p_N$.}
	\label{fig:nucleon_kinematics_0}
\end{figure}

\begin{figure}
	\centering
	\includegraphics[width=8cm]{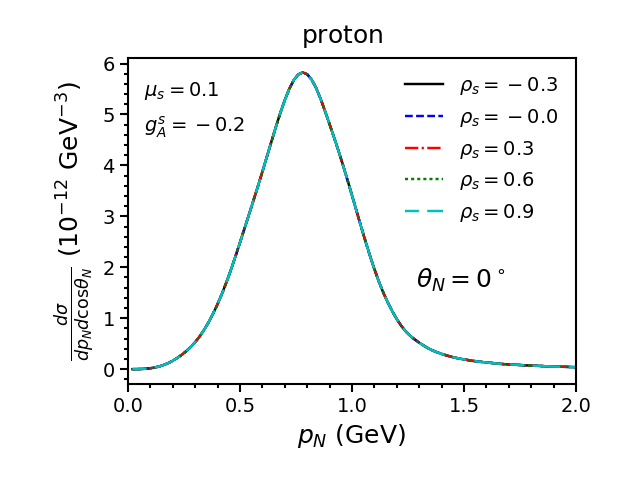}\\
	\includegraphics[width=8cm]{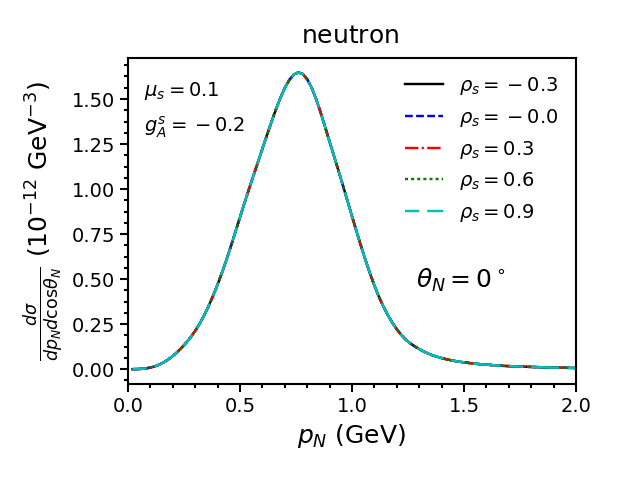}\\
	\includegraphics[width=8cm]{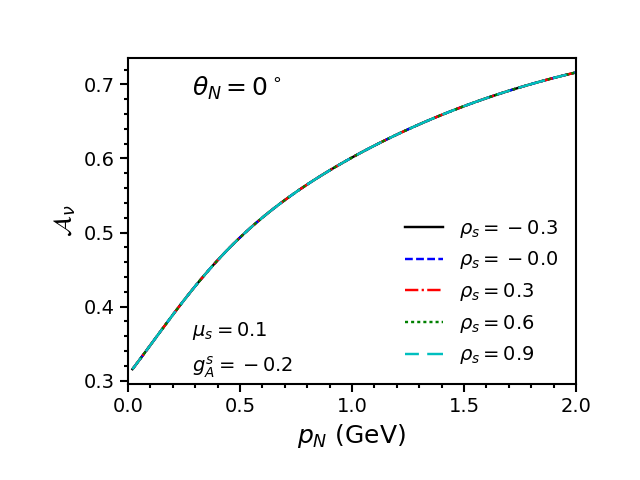} 	
	\caption{Neutral-current neutrino scattering from nucleons at $\theta_N=0^\circ$ as a function of the nucleon momentum $p_N$ for fixed $\mu_s$ and $g_A^s$ and varying values of $\rho_s$. The top panel shows the cross section for scattering from a proton; the middle panel shows the cross section for scattering from a neutron; the bottom panel shows the isospin asymmetry. }
	\label{fig:nucleon_0_rho}
\end{figure}

\begin{figure}
	\centering
	\includegraphics[width=8cm]{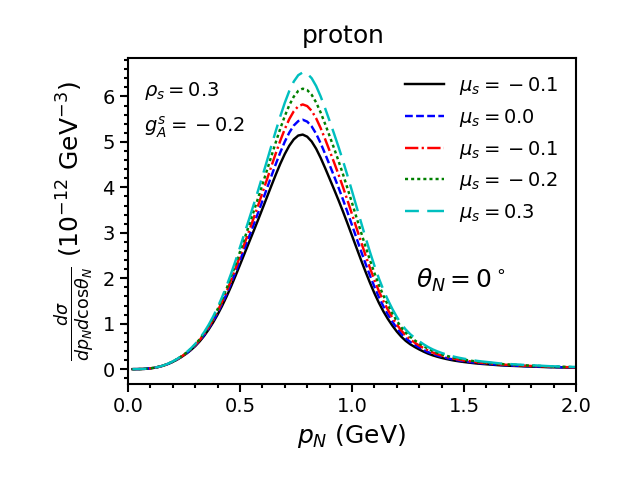}\\
	\includegraphics[width=8cm]{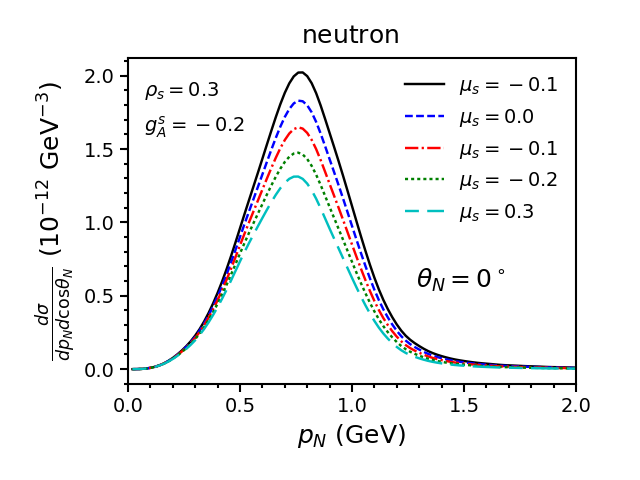}\\
	\includegraphics[width=8cm]{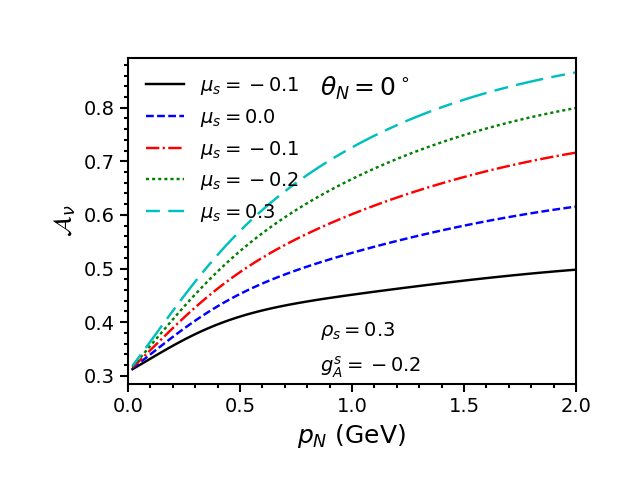} 	
	\caption{Same as Fig. \ref{fig:nucleon_0_rho}, but with $\rho_s$ and $g_A^s$ fixed and varying values of $\mu_s$.}
	\label{fig:nucleon_0_mu}
\end{figure}

\begin{figure}
	\centering
	\includegraphics[width=8cm]{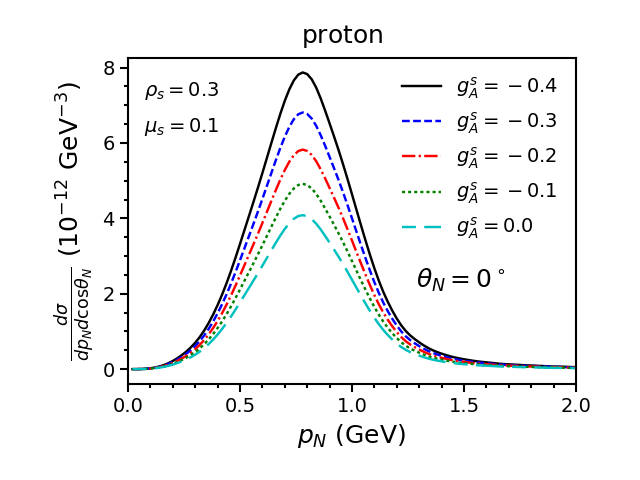}\\
	\includegraphics[width=8cm]{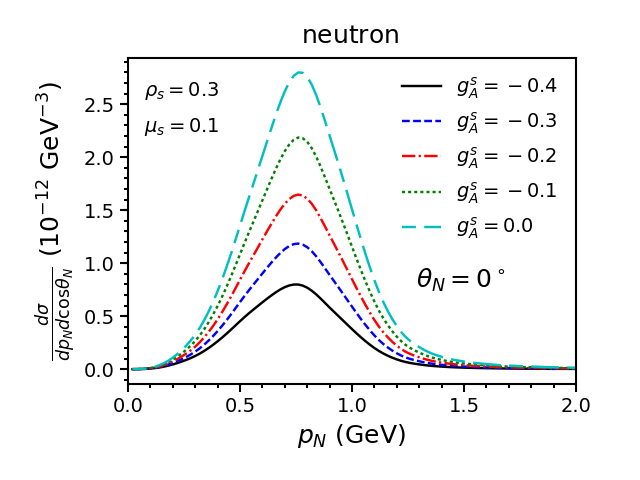}\\
	\includegraphics[width=8cm]{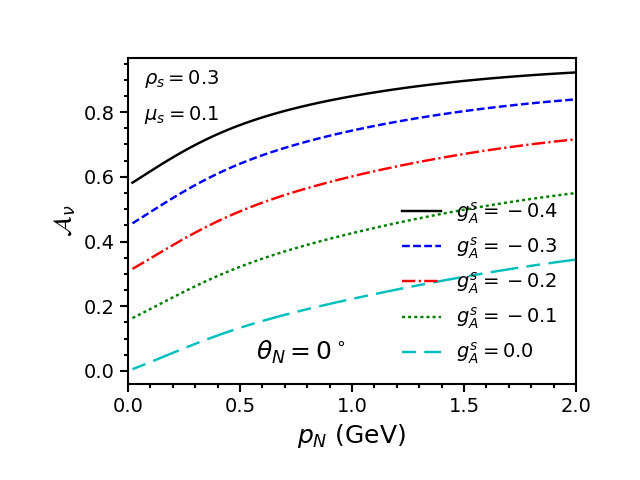} 	
	\caption{Same as Fig. \ref{fig:nucleon_0_rho}, but with $\rho_s$ and $\mu_s$ fixed and varying values of $g_A^s$.}
	\label{fig:nucleon_0_gas}
\end{figure}

We will first consider the case where the final-state three-momentum of the nucleon is at the angle $\theta_N=0^\circ$ to the neutrino beam. At this angle (\ref{eq:nucleon_0}) shows that the cross sections have no dependence on $\tilde{G}_E(\tau)$, but do depend on $\tilde{G}_M(\tau)$ and $\tilde{G}_A(\tau)$. Figure \ref{fig:nucleon_kinematics_0} shows the dependence on the neutrino energy of $k$ and the square on the four-momentum transfer $|Q^2|$ for these $\theta_N=0^\circ$ as a function of the magnitude of the final-state nucleon momentum $p_N$.  Figure \ref{fig:nucleon_0_rho} shows the proton (top panel) and neutron (center panel) cross sections along with the isospin asymmetry (bottom panel) for these kinematics, where $\mu_s$ and $g_A^s$ are fixed with varying $\rho_s$. As expected from the absence of $\tilde{G}_E(\tau)$ for this case the cross sections and asymmetry show no variation. Since the cross sections peak at about 0.75 GeV and fall to about 10 percent of the maximum value at about 1.2 GeV, possibly measurable values of $k$ range up to about 0.75 GeV, while the range of $|Q^2|$ is up to about 1 (GeV/c)$^2$. Figure \ref{fig:nucleon_0_mu} shows the cross sections and asymmetry for fixed $\rho_s$ and $g_A^s$ with varying $\mu_s$. The cross sections show moderate sensitivity to changes in the strangeness contribution to the magnetic form factor, while the asymmetry shows increasing sensitivity with increasing values of $p_N$. Figure \ref{fig:nucleon_0_gas} shows the cross sections and asymmetry for fixed $\rho_s$ and $\mu_s$ with varying $g_A^s$. The cross sections show substantial sensitivity to changes in the strangeness contribution to the axial-vector form factor and the asymmetry show substantial variation across the range of $p_N$.


\begin{figure}
	\centering
	\includegraphics[width=8cm]{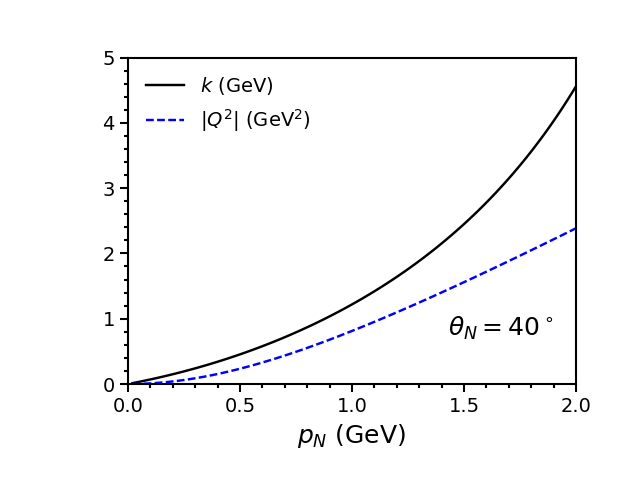} 	
	\caption{Values of $k$ and $|Q^2|$ for neutral-current neutrino scattering from a nucleon  at $\theta_N=40^\circ$ as a function of the nucleon momentum $p_N$.}
	\label{fig:nucleon_kinematics_40}
\end{figure}

\begin{figure}
	\centering
	\includegraphics[width=8cm]{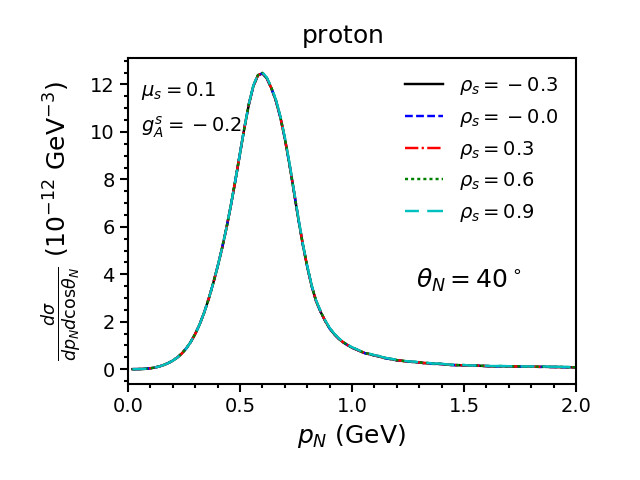}\\
	\includegraphics[width=8cm]{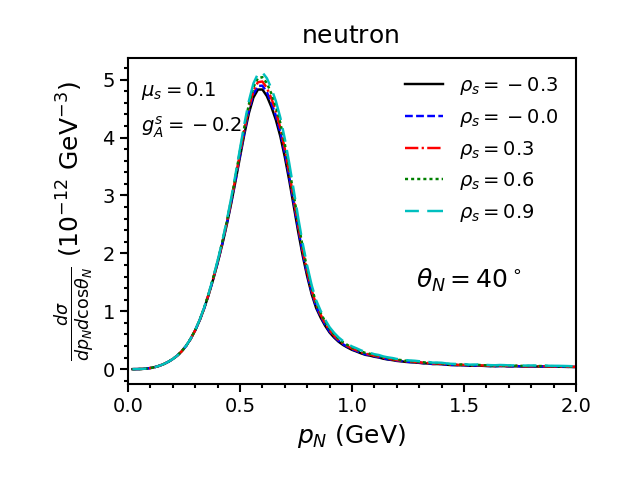}\\
	\includegraphics[width=8cm]{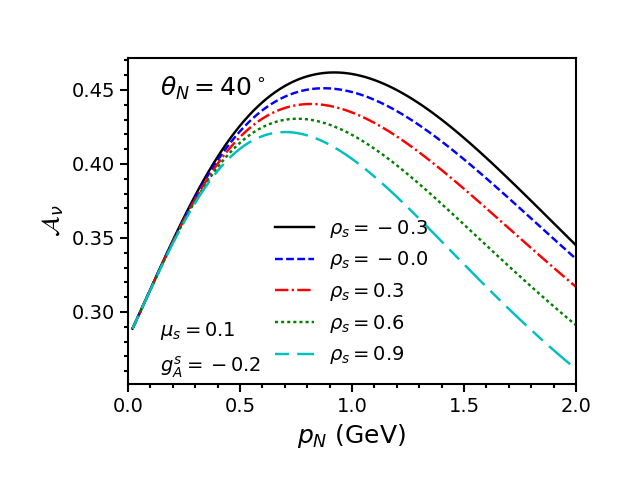} 	
	\caption{Neutral-current neutrino scattering at $\theta_N=40^\circ$ as a function of the nucleon momentum $p_N$ for fixed $\mu_s$ and $g_A^s$ with varying values of $\rho_s$. The top panel shows the cross section for scattering from a proton; the middle panel shows the cross section for scattering from a neutron; the bottom panel shows the isospin asymmetry.}
	\label{fig:nucleon_40_rho}
\end{figure}

\begin{figure}
	\centering
	\includegraphics[width=8cm]{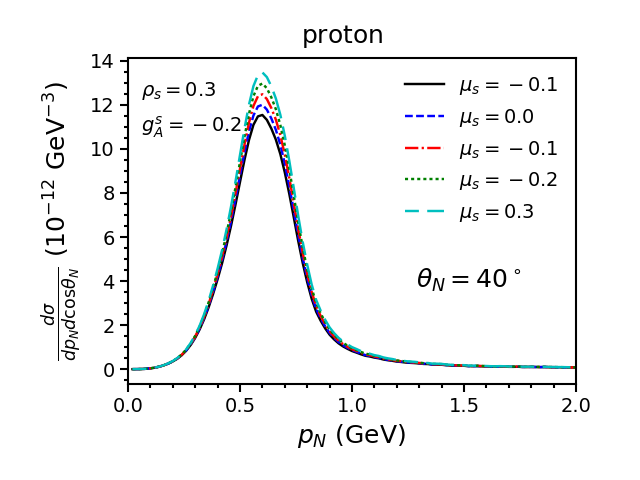}\\
	\includegraphics[width=8cm]{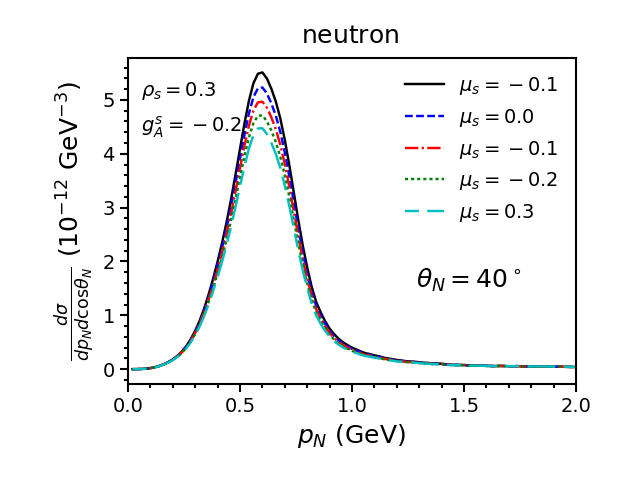}\\
	\includegraphics[width=8cm]{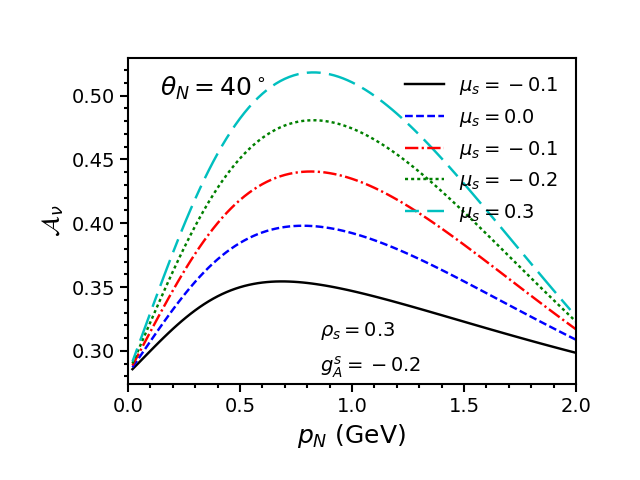} 	
	\caption{Same as Fig. \ref{fig:nucleon_40_rho}, but with $\rho_s$ and $g_A^s$ fixed and varying values of $\mu_s$.}
	\label{fig:nucleon_40_mu}
\end{figure}

\begin{figure}
	\centering
	\includegraphics[width=8cm]{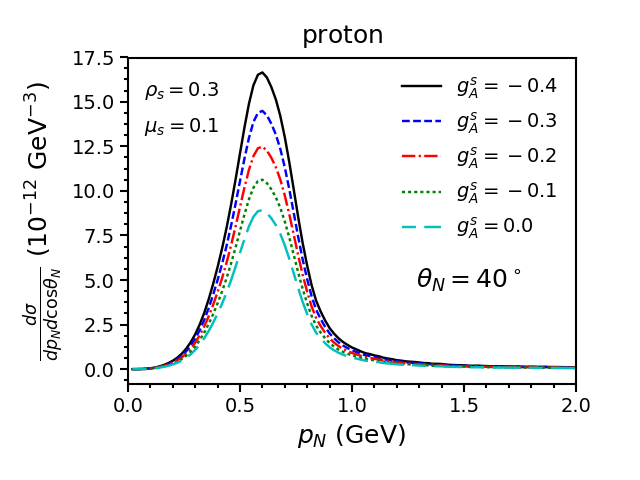}\\
	\includegraphics[width=8cm]{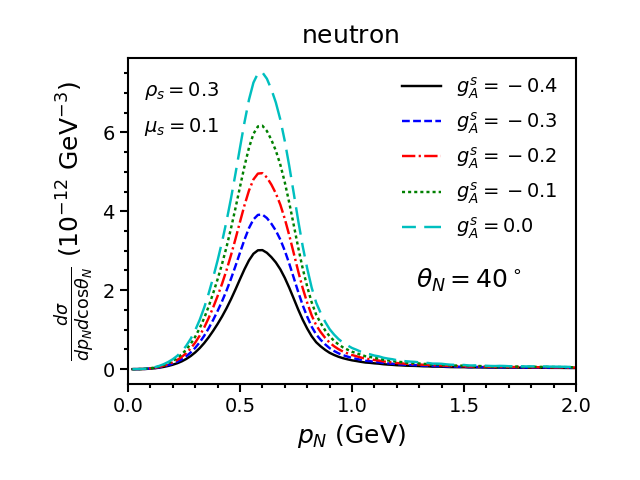}\\
	\includegraphics[width=8cm]{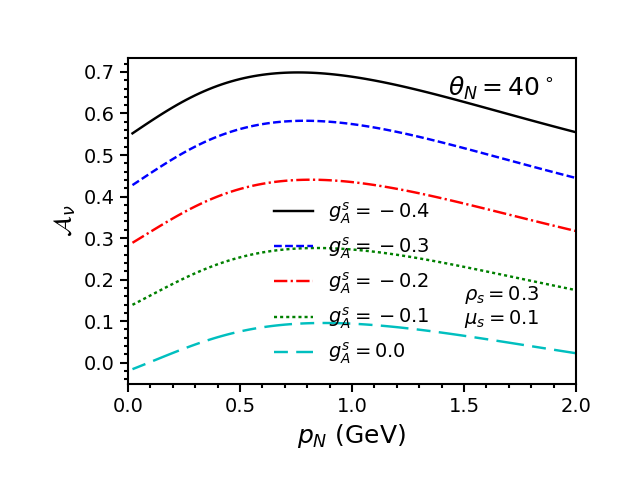} 	
	\caption{Same as Fig. \ref{fig:nucleon_40_rho}, but with $\rho_s$ and $\mu_s$ fixed and varying values of $g_A^s$.}
	\label{fig:nucleon_40_gas}
\end{figure}

In order to see any effect due to the strangeness contributions to the electric form factor, it is necessary to consider finite values of $\theta_N$. As an example we now present calculations for $\theta_N=40^\circ$. Figure \ref{fig:nucleon_kinematics_40} shows the values of $k$ and $|Q^2|$ for this choice of kinematics. Note that compared with Fig. \ref{fig:nucleon_kinematics_0}, the ranges of values for both of these quantities increase with $\theta_N$. This is particularly the case for $k$ due to the possibility that the denominator of (\ref{eq:k_single}) can vanish resulting in a kinematical singularity, which is also present in the cross section as can be seen from (\ref{eq:nucleon_cross_section}-\ref{eq:Upsilon_3}). Figure \ref{fig:nucleon_40_rho} shows the cross sections and asymmetry for $\theta_N=40^\circ$ with $\mu_s$ and $g_A^s$ fixed with varying $\rho_s$. The cross sections show only a small sensitivity due to the strangeness contributions to the electric form factor. The sensitivity is clearly enhanced in the isospin asymmetry, but is still small. The peak in the cross sections is shifted to slightly lower values of $p_N$ and the widths are also slightly increased. Figure \ref{fig:nucleon_40_mu} shows the cross sections and asymmetry for $\theta_N=40^\circ$ with $\rho_s$ and $g_A^s$ fixed, and with varying $\mu_s$. The sensitivity of the cross sections to the strangeness contribution to the magnetic form factor is still moderate, but is smaller than at $\theta_N=0^\circ$. Figure \ref{fig:nucleon_40_gas} shows the cross sections and asymmetry for $\theta_N=40^\circ$ with $\rho_s$ and $\mu_s$ fixed, and with varying $g_A^s$. The sensitivity to the strangeness contribution to the axial-vector form factor is still substantial even though slightly smaller than at $\theta_N=0^\circ$

Extension of the cross sections to larger $\theta_N$ will result in larger cross sections due to the kinematical singularity. The peaks of the cross sections move to lower values of $p_N$ and the widths also decrease. As a result, the range of $|Q^2|$ for extraction of the strangeness contributions to the form factors then decreases. This indicates that extraction of the strangeness contributions to the form factors for large $|Q^2|$  is most likely to be best done at forward angles. The sensitivity to strangeness contribution to the electric form factor is small while those to the magnetic and axial-vector form factors are moderate and large using this approach.

The problem with this exercise is that there are no practical free neutron targets that could be used in this reaction.  This means that neutron cross sections must be obtained by scattering from few- or many-body nuclei. The simplest of these is of course deuterium. In the following discussion we show that by measuring both the proton and neutron three-momenta for neutral-current disintegration of the deuteron the lepton kinematics for the reaction can be determined and that the deuteron can be used to produce an isospin asymmetry that has a sensitivity to the strangeness parameters comparable to the ideal, but unachievable single-nucleon case.

\section{Neutrino Scattering from the Deuteron}\label{sec:deuteron}

As there are no practical free neutron targets, the best alternative is to use a 
deuteron or helium-3 target.  Here we choose to focus on the deuteron, the simplest case. Since the deuteron is a bound state of a proton and neutron, it is clear that calculation of the isospin asymmetry, as done above for cross sections on free protons and neutrons, requires that kinematics can be found which isolate the neutron and proton contributions to the reaction with little interference between the two contributions. We will show that this is indeed possible. 

\begin{figure}[h]
	\centering
	\includegraphics[height=1.5in]{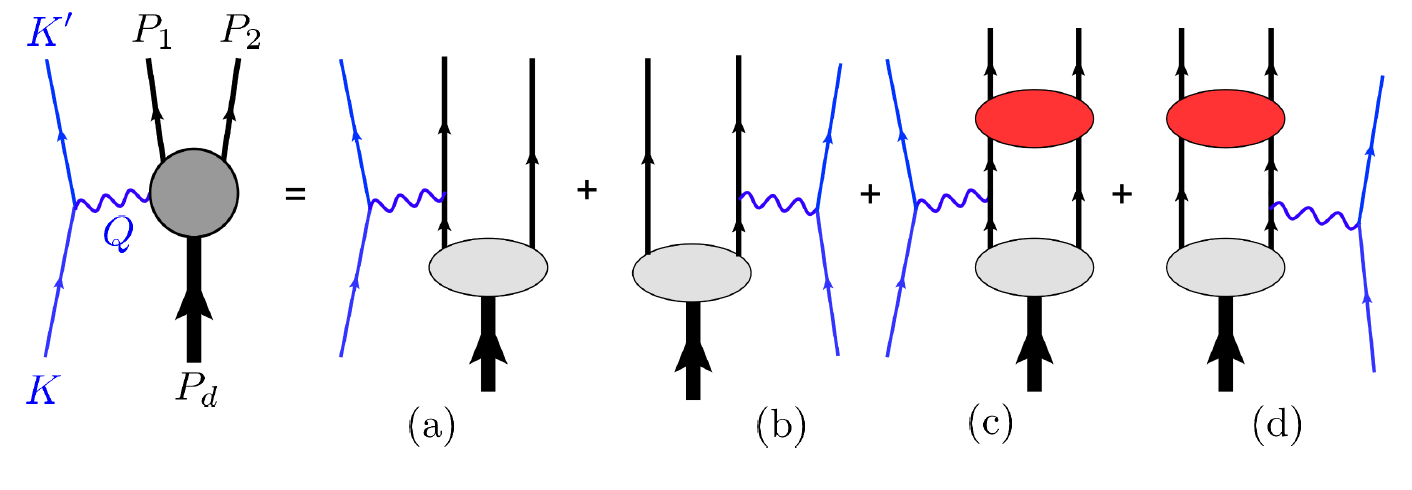} 
	\caption{Diagrams representing neutral-current neutrino scattering from a deuteron. Diagrams (a) and (b) are the plane-wave contributions to the reaction, while diagrams (c) and (d) are contributions from final-state interactions. The light ellipse represents the deuteron vertex function, while the dark ellipse represents the completely antisymmetric t matrix.}
	\label{fig:deuteron_diagrams}
\end{figure}

The calculations presented here are based on a model originally designed to examine deuteron electrodisintegration in kinematical regions similar to those occurring in long-baseline neutrino experiments where large energy and momentum transfers can occur making it necessary that Lorentz covariance be required. The model has been applied to electrodisintegration without polarization \cite{Jeschonnek:2008zg}, with polarized deuterons \cite{Jeschonnek:2009tq,Jeschonnek:2016jyn}, with polarized final-state protons \cite{Jeschonnek:2009ds} and to study the model dependence in these processes \cite{Ford:2014yua,Jeschonnek:2016jyn}. Feynman diagrams representing the elements of the model are given in Fig. \ref{fig:deuteron_diagrams}. The deuteron vertex function, represented as the light ellipse in Fig. \ref{fig:deuteron_diagrams} is obtained by a solution to the Gross equation \cite{Gross:1969rv,Gross:1972ye,Gross:1982nz} and is Lorentz covariant. Diagrams (a) and (b) represent neutral-current disintegration of the deuteron resulting in a free proton and neutron. Diagrams (c) and (d) include final-state interactions by means of an antisymmetrized proton-neutron t matrix represented by the dark ellipse. This final-state interaction is constructed either from helicity amplitudes taken from the SAID fit to nucleon-nucleon data \cite{SAIDdata} or from a Regge-model parameterization of NN scattering data from Mandelstam $s=5.4\ \mathrm{GeV^2}$ to $s=4000\ \mathrm{GeV^2}$ \cite{Ford:2012dg,Ford:2013zca,Ford:2013uza}. The nucleon with four-momentum $P_1$ is chosen to be the proton and that with $P_2$ is the neutron.  We have also extended the model to charged-current neutrino disintegration of the deuteron \cite{Moreno:2015nsa,VanOrden:2017uyy}.

The incident and final neutrino four-momenta are
\begin{equation}
K^\mu=(|\bm{k}|,\bm{k})
\end{equation}
and
\begin{equation}
{K'}^\mu=(|\bm{k}'|,\bm{k}')\,.
\end{equation}
The four-momentum transfer is
\begin{equation}
Q^\mu=K^\mu-{K'}^\mu=(\omega,\bm{q})\,.
\end{equation}
The final-state proton and neutron four-momenta are
\begin{equation}
P_1^\mu=(\sqrt{p_1^2+m_N^2},\bm{p}_1)
\end{equation}
and 
\begin{equation}
P_2^\mu=(\sqrt{p_2^2+m_N^2},\bm{p}_2)\,.
\end{equation}
The rest-frame four-momentum of the target deuteron is
\begin{equation}
P_d=(M_d,\bm{0})\,.
\end{equation}
 We have assumed that the neutrinos are massless.

\begin{figure}[h]
	\centering
	\includegraphics[width=8cm]{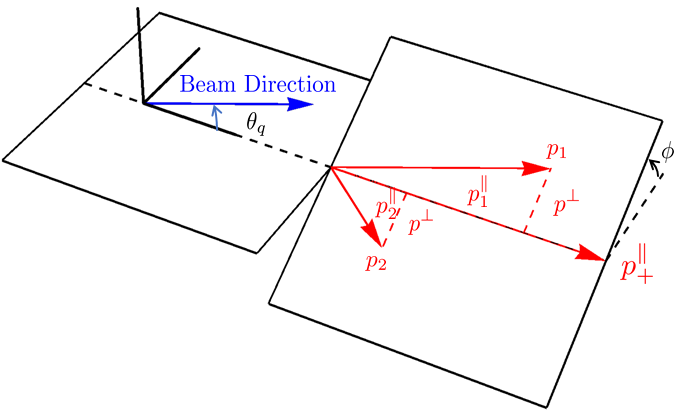} 
	\caption{Diagram representing the choice of kinematic variables for the $\nu+d\rightarrow\nu+p+n$ reaction.}
	\label{fig:deuteron_kinematics}
\end{figure}

Since kinematics associated with these diagrams are symmetric under the interchange of $P_1$ and $P_2$, it is convenient to express the semi-inclusive cross section in a form that manifests the same symmetry. The procedure for obtaining the necessary quantities to evaluate the cross section is as follows. Since the incident and scattered neutrinos cannot be detected, a complete set of variables that parameterize the cross section must be obtained by measuring the three-momentum of the proton $\bm{p}_1$ and of the neutron $\bm{p}_2$. These two momenta define the hadron plane for the reaction. Conservation of three-momentum gives
\begin{equation}
\bm{q}=\bm{p}_1+\bm{p}_2\equiv \bm{p}_+\,.
\end{equation} 
The momentum $\bm{q}$ and the beam direction define the lepton plane of the reaction and $\theta_q$ represents the angle between the beam direction and $\bm{q}$. Choosing the quantization axis $\bm{e}_z$ to be along $\bm{q}$, the azimuthal angle between the lepton and hadron planes is $\phi$. The proton and neutron momenta can then be decomposed into components parallel and perpendicular to $\bm{e}_z$ as
\begin{equation}
\bm{p}_1=\bm{p}_1^\perp+p_1^\parallel\bm{e}_z
\end{equation}
and
\begin{equation}
\bm{p}_2=\bm{p}_2^\perp+p_2^\parallel\bm{e}_z\,.
\end{equation}
Then
\begin{equation}
p_+^\parallel=p_1^\parallel+p_2^\parallel=q\,,
\end{equation}
and
\begin{equation}
\bm{p}_1^\perp+\bm{p}_2^\perp=0\,.
\end{equation}
So
\begin{equation}
|\bm{p}_1^\perp|=|\bm{p}_2^\perp|\equiv p_+^\perp
\end{equation}
with the azimuthal angles of the momenta being $\phi_1=\phi$ and $\phi_2=\phi+\pi$.
There are now five independent quantities that determine the cross section: $p_1^\parallel$, $p_2^\parallel$, $p^\perp$, $\theta_q$ and $\phi$. This procedure is illustrated by Fig. \ref{fig:deuteron_kinematics}. 


From conservation of energy
\begin{equation}
\omega+M_d=\sqrt{p_1^2+m_N^2}+\sqrt{p_2^2+m_N^2}=\sqrt{{p_1^\parallel}^2+{p^\perp}^2+m_N^2}+\sqrt{{p_2^\parallel}^2+{p^\perp}^2+m_N^2}\equiv E_+\,,
\end{equation}
or
\begin{equation}
\omega=E_+-M_d=k-k'\,.
\end{equation}
Now using
\begin{equation}
\bm{k}'=\bm{k}-\bm{q}\,,
\end{equation}
gives
\begin{equation}
k'=\sqrt{k^2-2kq\cos\theta_q+q^2}=\sqrt{k^2-2k p_+^\parallel\cos\theta_q+{p_+^\parallel}^2}=k-\omega=k-E_++M_d\,.
\end{equation}
This can be solved to give the initial neutrino energy
\begin{equation}
k=\frac{{p_+^\parallel}^2-(E_+-M_d)^2}{2(p_+^\parallel\cos\theta_q-E_++M_d)}=\frac{|Q^2|}{2(q\cos\theta_q-\omega)}\,.\label{eq:k}
\end{equation}
Note that this becomes singular at $\omega=q\cos\theta_q$.
The scattered neutrino energy is
\begin{equation}
k'=k-\omega=k-E_++M_d=\frac{{p_+^\parallel}^2-2(E_+-M_d)p_+^\parallel\cos\theta_q+(E_+-M_d)^2}{2(p_+^\parallel\cos\theta_q-E_++M_d)},.
\end{equation}
The neutrino scattering angle is given by
\begin{equation}
\cos\theta_l=\frac{k-p_+^\parallel\cos\theta_q}{k'}\,.
\end{equation}

Using these coordinates the semi-inclusive cross section can be written as
\begin{equation}
\left(\frac{d\sigma}{dp^\perp dp_1^\parallel dp_2^\parallel 
	d\phi}\right)_\chi
=\frac{G^2\cos^2\theta_cm_N^2 p^\perp}{2(2\pi)^5 
	E_1E_2(p_+^\parallel\cos\theta_q-E_++M_d)}
\frac{P(k)}{k}v_0
\mathcal{F}^2_\chi\,,
\end{equation}
where
\begin{equation}
E_i=\sqrt{{p_i^\parallel}^2+{p^\perp}^2+m_N^2}
\end{equation}
and
\begin{align}
{\cal F}_\chi^2=&\hat{V}_{CC}
(w_{CC}^{VV(I)}+w_{CC}^{AA(I)})
+2\hat{V}_{CL}
(w_{CL}^{VV(I)}+w_{CL}^{AA(I)})
+\hat{V}_{LL}
(w_{LL}^{VV(I)}+w_{LL}^{AA(I)})\nonumber\\
&+\hat{V}_T
(w_{T}^{VV(I)}+w_{T}^{AA(I)})\nonumber\\
&+\hat{V}_{TT}
\left[(w_{TT}^{VV(I)}+w_{TT}^{AA(I)})
\cos 2\phi+
(w_{TT}^{VV(II)}+w_{TT}^{AA(II)})
\sin 2\phi\right]\nonumber\\
&+\hat{V}_{TC}
\left[(w_{TC}^{VV(I)}+w_{TC}^{AA(I)})
\cos\phi
+(w_{TC}^{VV(II)}+w_{TC}^{AA(II)})
\sin\phi)\right]\nonumber\\
&+\hat{V}_{TL}
\left[(w_{TL}^{VV(I)}+w_{TL}^{AA(I)})
\cos\phi
+(w_{TL}^{VV(II)}+w_{TL}^{AA(II)})
\sin\phi\right]
\nonumber\\
&+\chi\left[\hat{V}_{T'}
w^{VA(I)}_{T'}+
\hat{V}_{TC'}(w^{VA(I)}_{TC'}
\cos\phi
+w_{TC'}^{VA(II)}
	\sin\phi)\right.\nonumber\\
&\left.+\hat{V}_{TL'}
(w^{VA(I)}_{TL'}
\cos\phi
+w^{VA(II)}_{TL'}
	\sin\phi)\right]\,.
\end{align}
The kinematic factors are given by (\ref{eq:kinematic_factors}) and the response functions $w_i^j$ are defined in \cite{Moreno:2014kia}.  This expression is in a form that easily allows the interchange $1\leftrightarrow2$. 

For the purposes of this work it is convenient to introduce the variable
\begin{equation}
y=\frac{(M_d+\omega)\sqrt{s(s-4m_N^2)}}{2s}
-\frac{q}{2}=\frac{E_+\sqrt{(E_+^2-q^2)(E_+^2-{p_+^\parallel}^2-4m_N^2)}}{2(E_+^2-{p_+^\parallel}^2)}-\frac{p_+^\parallel}{2} \,,
\end{equation}
since the deuteron cross section will tend to be maximal for $y=0$. 
For $y=0$ this can be solved to yield
\begin{align}
p_+^\parallel=\sqrt{E_+(E_+-2m_N)}\,.
\end{align}
This, in turn can be solved for $p^\perp$ to give
\begin{equation}
p^\perp=\sqrt{\frac{2
		m_N p_1^\parallel p_2^\parallel
		\left(\sqrt{m_N^2+(p_1^\parallel+p_2^
			\parallel)^2}-m_N\right)}{(p_1^\parallel
		+p_2^\parallel)^2}} \,.
\end{equation}
This constraint can be used to reduce the number of independent parameters to four.

The four-momentum transfer must be space-like,  $|Q^2|\geq 0$. From (\ref{eq:k}) it is clear that $k$ becomes singular at $q\cos\theta_q=\omega$. A  constraint that avoids this singularity and keeps the four-momentum transfer space-like can be roughly enforced by requiring that $k<k_\mathrm{max}$, where $k_\mathrm{max}$ is chosen from an examination of the neutrino flux factor.  An approximate constraint on the input parameters is then given by
\begin{equation}
p_2^\parallel+p_1^\parallel<\frac{2	k_\mathrm{max} m_N	(k_\mathrm{max}+m_N) \cos	\theta_q}{k_\mathrm{max}^2 \sin^2\theta_q+2 k_\mathrm{max}	m_N+m_N^2}\,.
\end{equation}
For the calculations shown in this paper we have chosen to use a flux factor related to the DUNE flux and have chosen $k_\mathrm{max}=10\ \mathrm{GeV}$.

\begin{figure}
	\centering
\includegraphics[width=8cm]{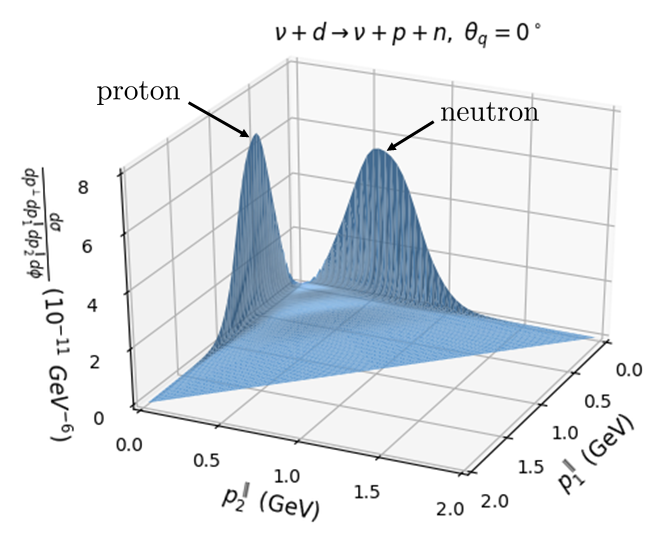} 	
\caption{The cross section for neutral-current scattering from the deuteron for $\theta_q=0^\circ$ as a function of $p^\parallel_1$ and $p^\parallel_2$. Regions of the cross section which are primarily due to scattering from a proton or neutron are indicated.}
\label{fig:deuteron_3d_fsi}
\end{figure}

The question of whether or not it is possible to separate the neutron and proton contributions to this cross section can be answered by considering Fig. \ref{fig:deuteron_3d_fsi}. Here, the semi-inclusive cross section is plotted for $\theta_q=0^\circ$ as a function of $p_1^\parallel$ and $p_2^\parallel$. This includes all of the contributions from the diagrams of Fig. \ref{fig:deuteron_diagrams} and therefore includes final-state interactions (FSI). This figure shows two well-separated peaks that can be identified as contributions arising primarily from scattering from the proton and the neutron.  By fixing $p_2^\parallel$ at a small value and allowing $p_1^\parallel$ to run from zero to 2 GeV, the cross section from the proton can be extracted. Conversely, fixing $p_1^\parallel$ at a small value and allowing $p_2^\parallel$ to vary it is possible to extract the contribution from the neutron. This then makes it possible to construct the isospin asymmetry   
\begin{equation}
	\mathcal{A}_\chi=\frac{ \left(d\sigma/dp^\perp dp_1^\parallel dp_2^\parallel 
			d\phi\right)_\chi^p- 
		\left(d\sigma/dp^\perp dp_1^\parallel dp_2^\parallel 
			d\phi\right)_\chi^n}{ \left(d\sigma/dp^\perp dp_1^\parallel dp_2^\parallel 
			d\phi\right)_\chi^p+ \left(d\sigma/dp^\perp dp_1^\parallel dp_2^\parallel 
			d\phi\right)_\chi^n}
\end{equation}
from the deuteron cross sections.

\subsection{Results: Sensitivity of the Differential Cross Sections and Asymmetries to the Strangeness Parameters}

We can now proceed in the same manner as for the free nucleon case. For the results presented here we will set $p_2^\parallel(p_1^\parallel)=0.001\ \mathrm{GeV}$ and define $p^\parallel=p_1^\parallel(p_2^\parallel)$ for the proton(neutron) cross sections. This choice maximizes the size of the cross sections.

\begin{figure}
	\centering
	\includegraphics[width=8cm]{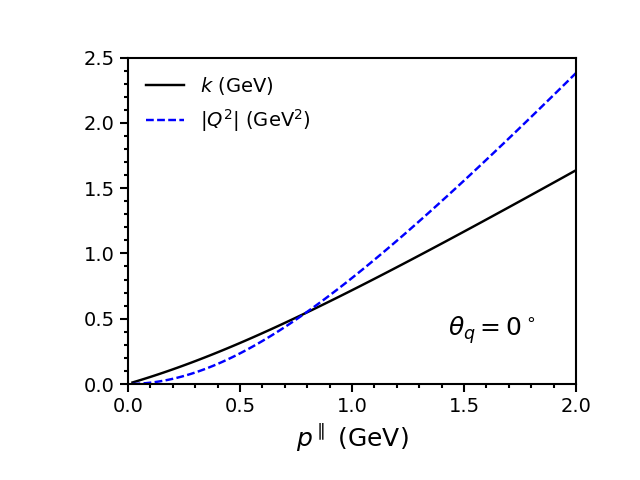}\\
	\includegraphics[width=8cm]{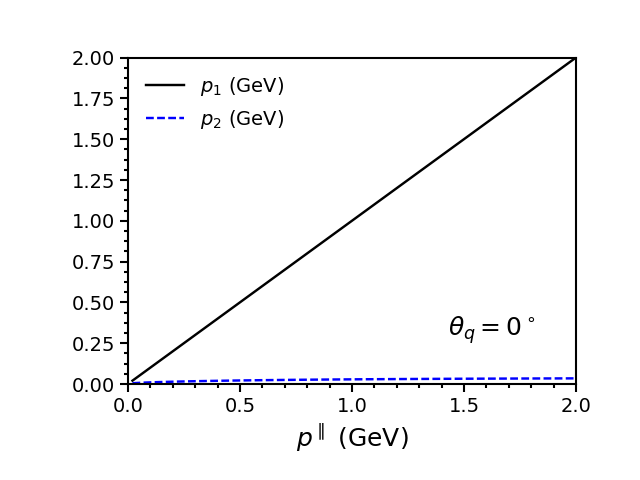}\\
	\includegraphics[width=8cm]{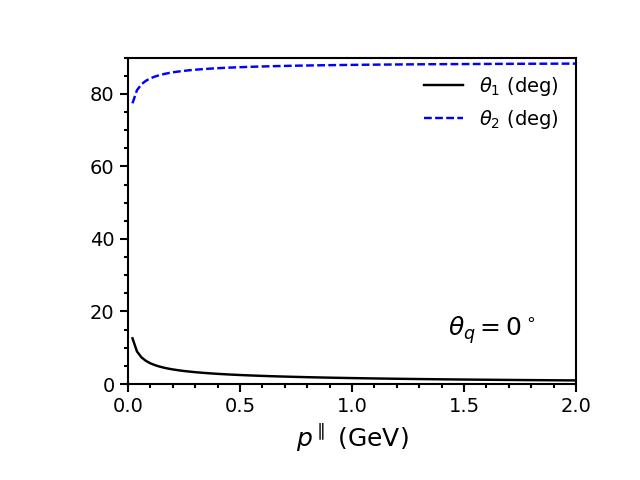} 	
	\caption{Kinematics for neutral-current neutrino scattering from deuterium at $\theta_q=0^\circ$ as a function of the parallel component of $p_1(p_2)=p^\parallel$. The top panel shows $k$ and $|Q^2|$; the center panel shows $p_1$ and $p_2$; the bottom panel shows the nucleon angles $\theta_1$ and $\theta_2$ relative to $\bm{q}$.}
	\label{fig:deuteron_kinematics_0}
\end{figure}

Figure \ref{fig:deuteron_kinematics_0} gives the kinematics for $\theta_q=0^\circ$. The top panel shows $k$ and $|Q^2|$ as a function of $p^\parallel$. Comparison of this figure with Fig. \ref{fig:nucleon_kinematics_0} shows that these quantities are essentially the same as for scattering from a nucleon at $\theta_N=0^\circ$. The center panel shows the magnitude of the three momenta for the proton kinematics. Note that, in the case of scattering from a free proton, $p_N$ is equivalent to $p_1$ and $p_2$ is not defined. The bottom figure shows the angles of the two nucleons with respect to $\bm{q}$ as a function of $p^\parallel$. Here $\theta_1$ begins at $13^\circ$ and decreases toward $0^\circ$ with increasing $p^\parallel$. The corresponding angle for scattering from a proton is $0^\circ$ for all values of $p^\parallel$.

\begin{figure}
	\centering
	\includegraphics[width=8cm]{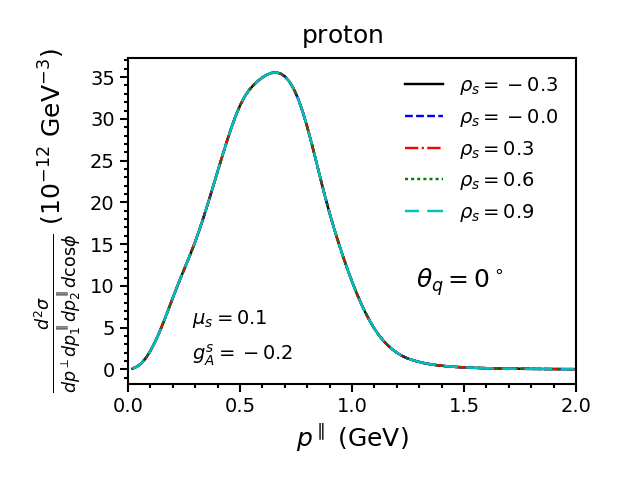}\\
	\includegraphics[width=8cm]{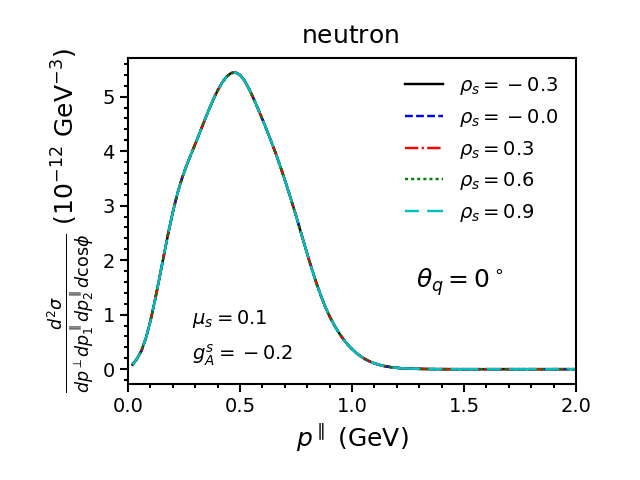}\\
	\includegraphics[width=8cm]{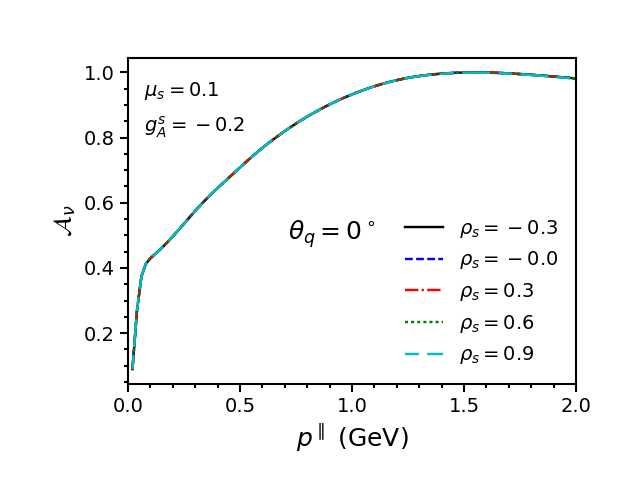} 	
	\caption{Neutral-current neutrino scattering from deuterium at $\theta_q=0^\circ$ as a function of  $p^\parallel$ for fixed $\mu_s$ and $g_A^s$ and varying values of $\rho_s$. The top panel shows the cross section for scattering from a proton; the middle panel shows the cross section for scattering from a neutron; the bottom panel shows the isospin asymmetry.}
	\label{fig:deuteron_0_rho}
\end{figure}

\begin{figure}
	\centering
	\includegraphics[width=8cm]{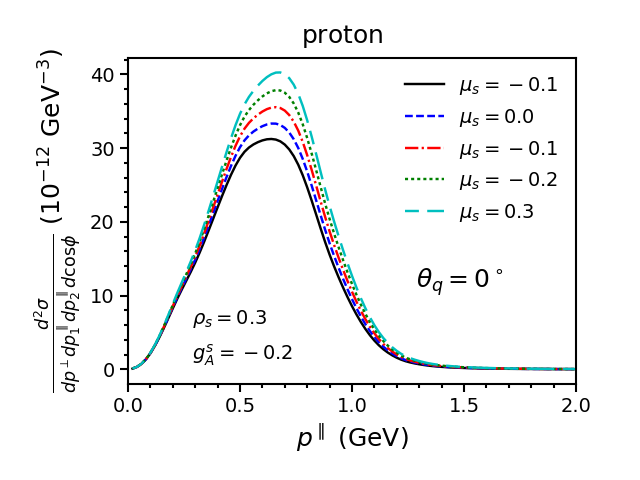}\\
	\includegraphics[width=8cm]{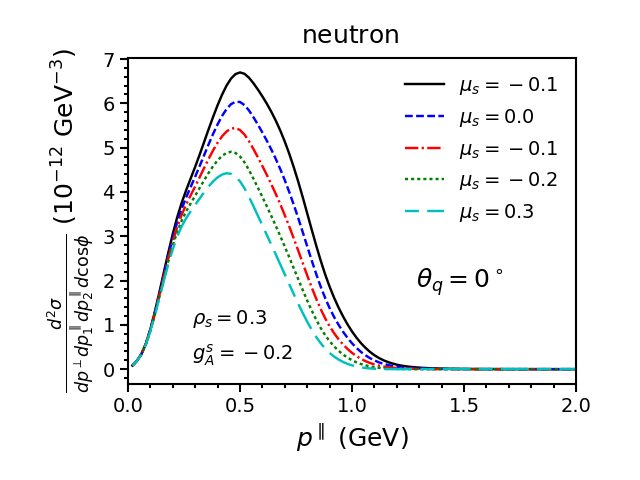}\\
	\includegraphics[width=8cm]{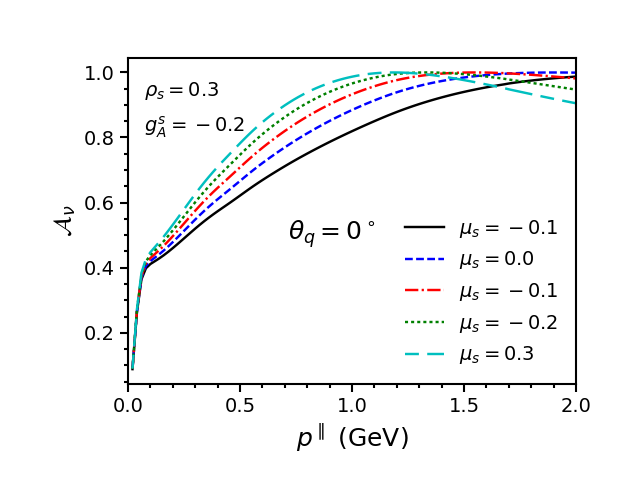} 	
	\caption{Same as Fig. \ref{fig:deuteron_0_rho}, but with $\rho_s$ and $g_A^s$ fixed and varying values of $\mu_s$.}
	\label{fig:deuteron_0_mu}
\end{figure}

\begin{figure}
	\centering
	\includegraphics[width=8cm]{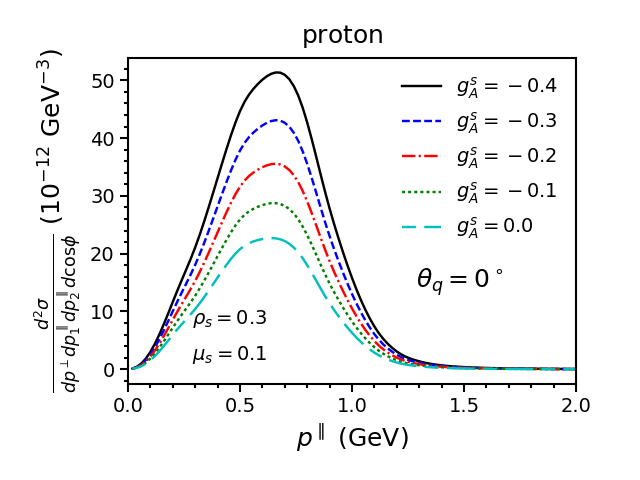}\\
	\includegraphics[width=8cm]{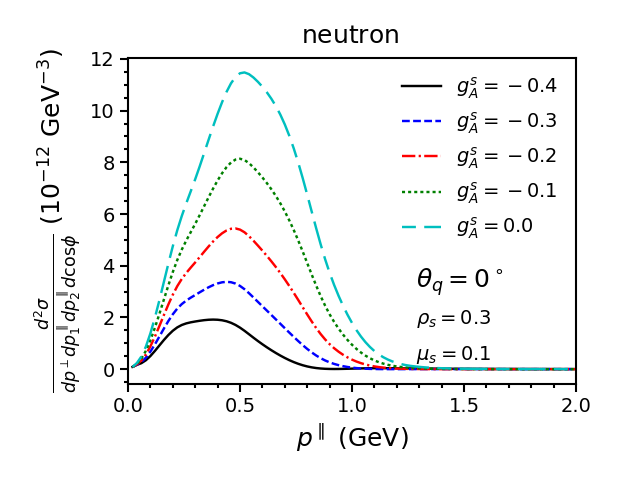}\\
	\includegraphics[width=8cm]{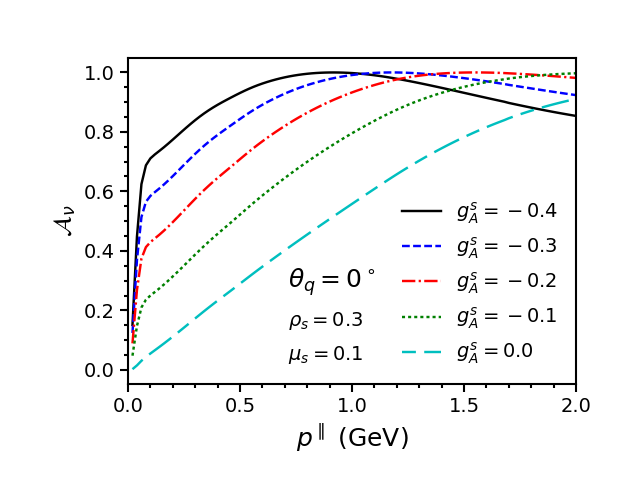} 	
	\caption{Same as Fig. \ref{fig:deuteron_0_rho}, but with $\rho_s$ and $\mu_s$ fixed and varying values of $g_A^s$.}
	\label{fig:deuteron_0_gas}
\end{figure}

Figure \ref{fig:deuteron_0_rho} gives the cross sections and isospin asymmetry for these kinematics with fixed $\mu_s$ and $g_A^s$ and varying $\rho_s$. The cross sections and asymmetry display no sensitivity to variation of $\rho_s$. Comparison with the free nucleon case of Fig. \ref{fig:nucleon_0_rho} shows that the peaks in the cross sections are moved to lower value of $p^\parallel$  and the shapes of the peaks and isospin asymmetry are different.  This is due in part to the factor of $p^\perp$ in the cross sections which vanishes rapidly as $p^\parallel$ goes to 0. Figure \ref{fig:deuteron_0_mu} gives the cross sections and isospin asymmetry for the kinematics with fixed $\rho_s$ and $g_A^s$ with varying $\mu_s$. As in the free nucleon case, the deuteron cross sections and asymmetry display moderate sensitivity to variation of the strangeness component of the magnetic form factor, but is slightly smaller in size. Figure \ref{fig:deuteron_0_gas} gives the cross sections and isospin asymmetry for the kinematics with fixed $\rho_s$ and $\mu_s$ with varying $g_A^s$. The cross sections and asymmetry display a large sensitivity to variations in the strangeness component of the axial-vector form factor. The asymmetry, however, becomes markedly less sensitive for the lowest values of $g_A^s$ for values of $p^\parallel$ above the peaks of the cross sections.

Although Figs. \ref{fig:deuteron_0_rho}, \ref{fig:deuteron_0_mu} and \ref{fig:deuteron_0_gas} are calculated including the FSI represented by Fig. \ref{fig:deuteron_diagrams} (c) and (d), for the kinematics shown here, the FSI contributions are small. This may not be the case for other kinematic choices. If the need arises, the model can be improved to provide a more accurate and consistent treatment of the FSI.

\begin{figure}
	\centering
	\includegraphics[width=8cm]{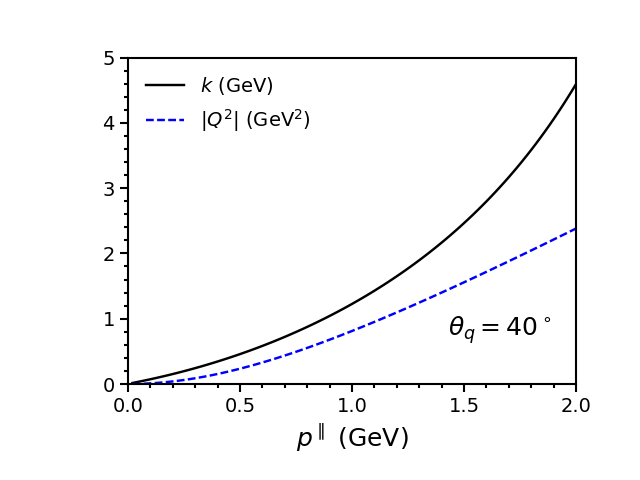}\\
	\includegraphics[width=8cm]{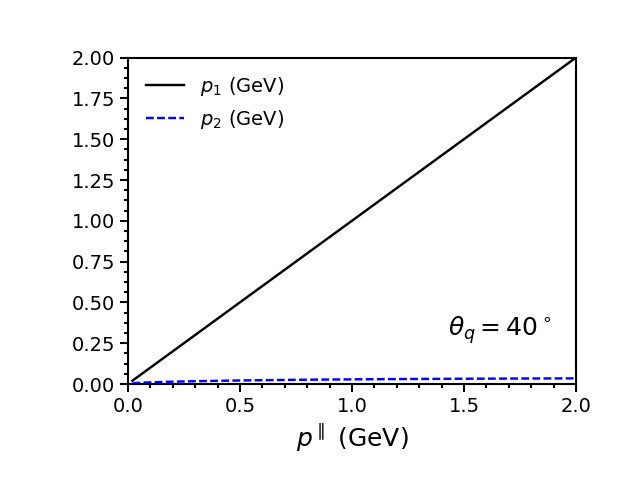}\\
	\includegraphics[width=8cm]{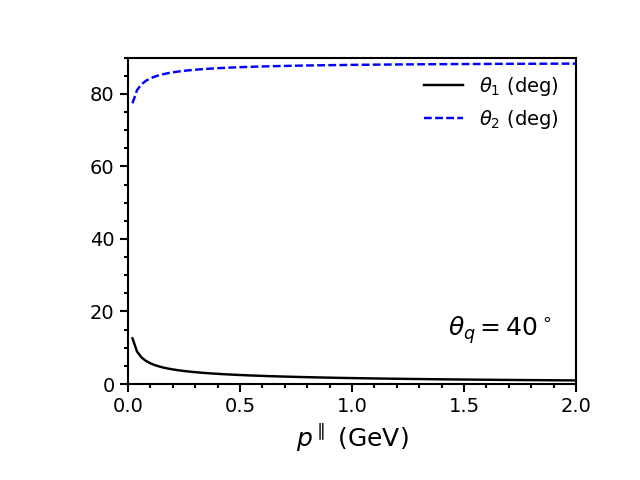} 	
	\caption{Kinematics for neutral-current neutrino scattering from deuterium at $\theta_q=40^\circ$ as a function of the parallel component of $p_1(p_2)=p^\parallel$. The top panel shows $k$ and $|Q^2|$; the center panel shows $p_1$ and $p_2$; the bottom panel shows the nucleon angles $\theta_1$ and $\theta_2$ relative to $\bm{q}$.}
	\label{fig:deuteron_kinematics_40}
\end{figure}
\begin{figure}
	\centering
	\includegraphics[width=8cm]{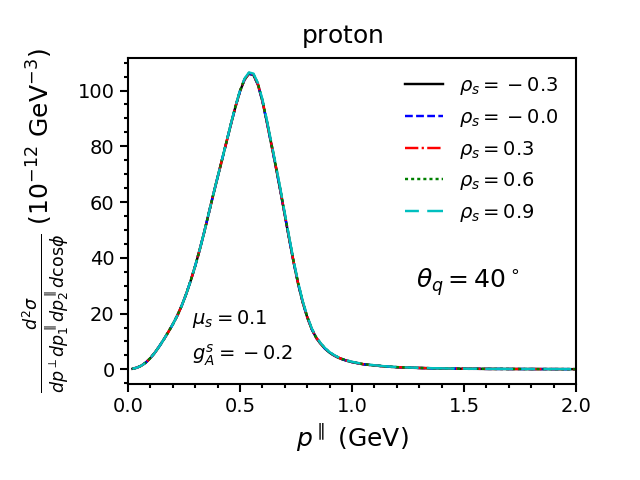}\\
	\includegraphics[width=8cm]{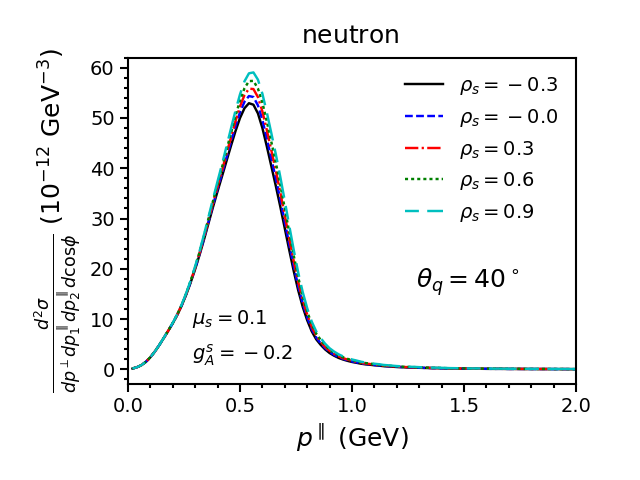}\\
	\includegraphics[width=8cm]{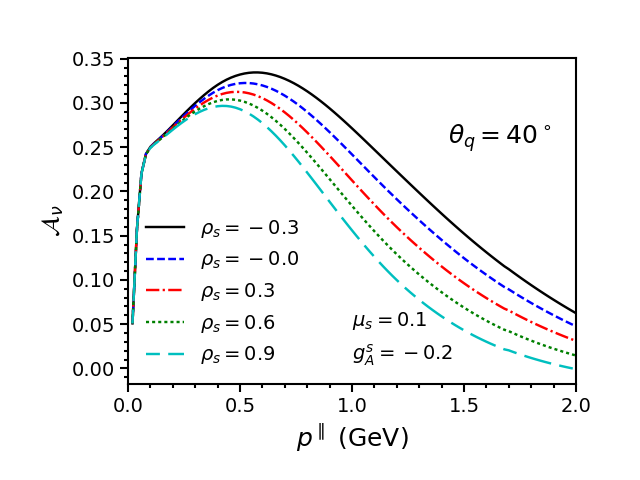} 	
	\caption{Neutral-current neutrino scattering from deuterium at $\theta_q=40^\circ$ as a function of  $p^\parallel$ for fixed $\mu_s$ and $g_A^s$ and varying values of $\rho_s$. The top panel shows the cross section for scattering from a proton; the middle panel shows the cross section for scattering from a neutron; the bottom panel shows the isospin asymmetry.}
	\label{fig:deuteron_40_rho}
\end{figure}
\begin{figure}
	\centering
	\includegraphics[width=8cm]{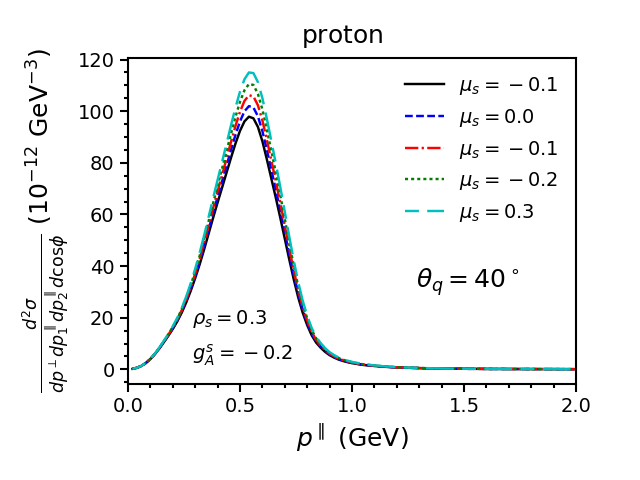}\\
	\includegraphics[width=8cm]{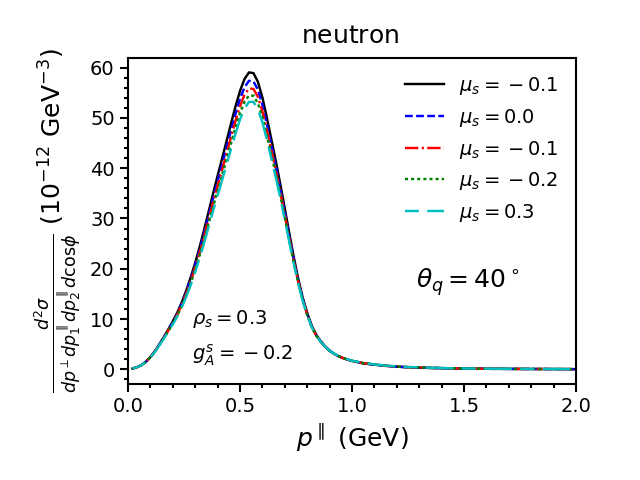}\\
	\includegraphics[width=8cm]{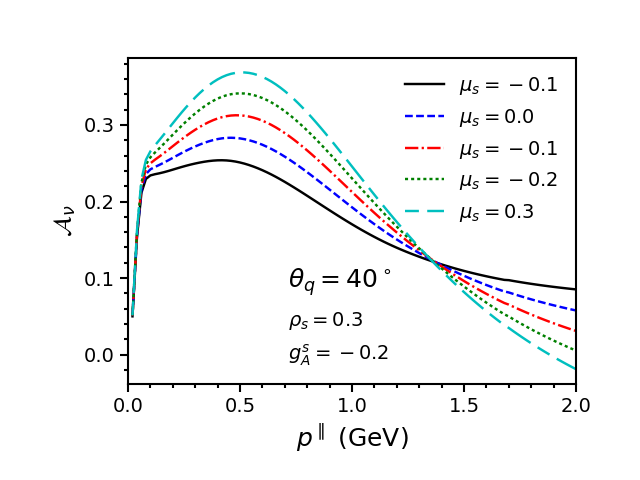} 	
	\caption{Same as Fig. \ref{fig:deuteron_40_rho}, but with $\rho_s$ and $g_A^s$ fixed and varying values of $\mu_s$.}
	\label{fig:deuteron_40_mu}
\end{figure}

Now consider the case where $\theta_q=40^\circ$. Figure \ref{fig:deuteron_kinematics_40} shows the kinematical variables for this case as a function of $p^\parallel$. Note that values of $k$ and $|Q^2|$ are almost identical to those for the free nucleons at $\theta_N=40^\circ$. The values of $p_1$ are virtually the same as $p^\parallel$ and $p_N$. The value of the recoil momentum $p_2$ is small. Again, $\theta_1$ starts at about $13^\circ$ and approaches $0^\circ$ with increasing $p^\parallel$. The recoil angle $\theta_2$ is near $90^\circ$. Figure \ref{fig:deuteron_40_rho} gives the cross sections and isospin asymmetry for the kinematics with fixed $\mu_s$ and $g_A^s$ and varying $\rho_s$. The cross sections and asymmetry now display a small sensitivity to variation of $\rho_s$. Comparison to the free nucleon case of Fig. \ref{fig:nucleon_0_rho} shows that the peaks in the cross sections are moved to lower value of $p^\parallel$  and the shapes of the peaks and isospin asymmetry are different.  Figure \ref{fig:deuteron_40_mu} gives the cross sections and isospin asymmetry for the kinematics with fixed $\rho_s$ and $g_A^s$ with varying $\mu_s$. As in the free nucleon case, the cross sections and asymmetry display moderate sensitivity to variation of the strangeness component of the magnetic form factor, but is slightly smaller in size. Figure \ref{fig:deuteron_0_gas} gives the cross sections and isospin asymmetry for the kinematics with fixed $\rho_s$ and $\mu_s$ with varying $g_A^s$. The cross sections and asymmetry display a large sensitivity to variations in the strangeness component of the axial-vector form factor. As in the case of $\theta_q=0^\circ$, the asymmetry, however, becomes markedly less sensitive for the lowest values of $g_A^s$ for values of $p^\parallel$ above the peaks of the cross sections.

\begin{figure}
	\centering
	\includegraphics[width=8cm]{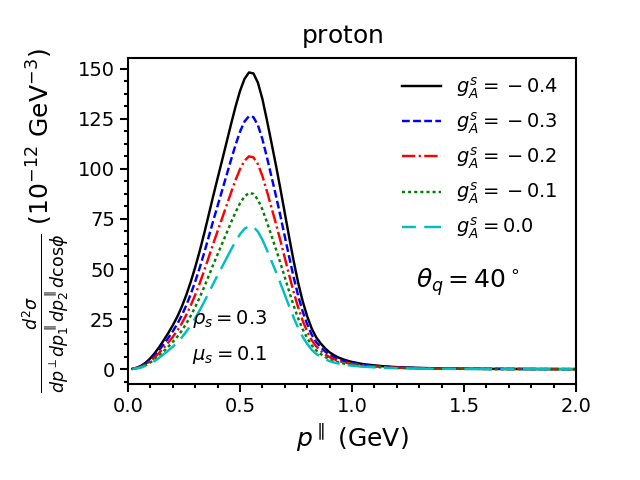}\\
	\includegraphics[width=8cm]{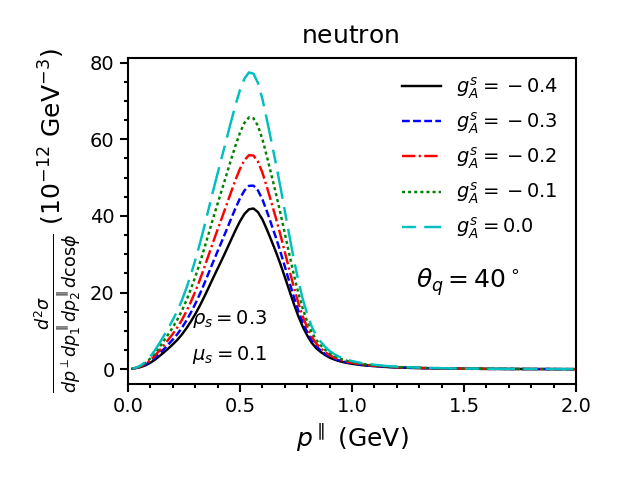}\\
	\includegraphics[width=8cm]{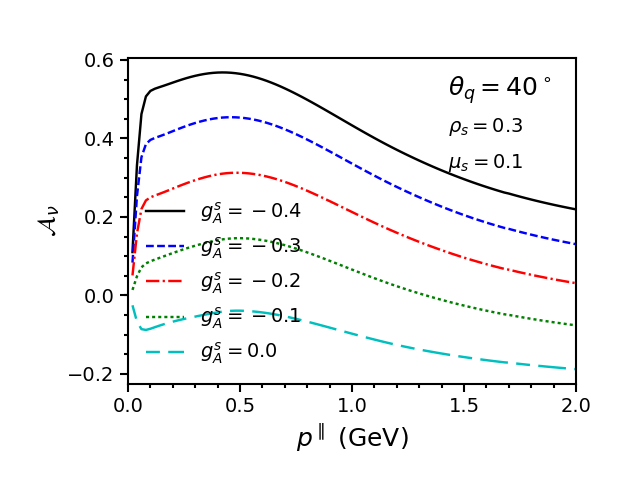} 	
	\caption{Same as Fig. \ref{fig:deuteron_40_rho}, but with $\rho_s$ and $\mu_s$ fixed and varying values of $g_A^s$.}
	\label{fig:deuteron_40_gas}
\end{figure}

Extension of the cross sections to larger $\theta_q$ is similar to the kinematical singularity as is the case for scattering from free nucleons. The range in $|Q^2|$ is significantly reduced while the size of the cross sections is increased. 

In the examples shown here we have chosen a very small value of the recoil momentum by setting $p_2^\parallel(p_1^\parallel)=1\ \mathrm{MeV}$ for protons(neutrons). This can be increased by choosing larger values for this parallel component of the recoil momentum. Increasing this value to $25\ \mathrm{MeV}$ will result in recoil momenta in the range of $100\ \mathrm{MeV}$. This is done at the expense of decreasing the size of the cross sections and introducing a region where the proton and neutron contributions overlap. 

In summary, the differential cross sections are practically insensitive to $\rho_s$, display a moderate sensitivity to $\mu_s$ and a fairly
strong sensitivity to $g_A^s$. The neutrino - proton and neutrino - neutron cross sections show sensitivity to $\mu_s$ and $g_A^s$ in the
opposite directions, so we considered asymmetries that exploit this behavior.

In the asymmetries, the sensitivity to $g_A^s$ appears to be quite strong, and measurements of the differential cross sections and asymmetries together at a handful of angles $\theta_q$
might well allow one to constrain the values for $\mu_s$  and $g_A^s$ better than methods that have been used up to this time.

\section{Summary and Outlook}\label{sec:summary}

We have presented theoretical results for neutral-current neutrino scattering from the deuteron leading to final hadronic states containing only a proton and neutron. In this work the focus has been placed on the fully exclusive reaction where both the proton and neutron in the final state are assumed to be detected. As discussed when considering the kinematics in detail, the incident neutrino energy, the neutrino scattering angle and the scattered neutrino’s energy can all be reconstructed given this information. 

The model employed is Lorentz covariant although the treatment of the FSI is not completely consistent with the calculation of deuteron vertex functions. This inconsistency can be removed with a moderate amount of effort. Since our interest is in modeling reactions where the typical neutrino energies are in the several GeV region, the ability to follow an approach where relativistic dynamics are incorporated is especially important. We note in passing that the close similarity between electrodisintegration of deuterium (for instance, via the $(e,e'p)$ reaction), charge-changing neutrino reactions on deuterium (namely, the CC$\nu$ reactions that can occur below pion production threshold, which are $\nu_\ell + ^2$H$ \rightarrow \ell^- + p + n$ and ${\bar\nu}_\ell + ^2$H$ \rightarrow \ell^+ + p + n$) and the NC$\nu$ reactions that provide the focus of the present work,  $\nu_\ell + ^2$H$\rightarrow \nu_\ell + p + n$ and ${\bar\nu}_\ell + ^2$H$\rightarrow {\bar\nu}_\ell + p + n$, ultimately should allow one to reach levels of theoretical uncertainty from the underlying nuclear physics to levels that are extremely hard to achieve otherwise.

We have shown that for specific choices of kinematics the reaction proceeds mainly via scattering from the proton, while for other choices of kinematics it proceeds mainly from the neutron. One can see this by doing as we have in this work and comparing the cross section for NC$\nu$ disintegration of the deuteron in these special kinematical situations with elastic neutrino scattering from a proton or neutron. Thus a special feature of such studies using the deuteron as a target is to have both protons and neutrons as effective targets for neutral-current neutrino scattering with well-determined kinematics for the leptonic part of the problem. Given this ability to isolate effects from protons and neutrons then provides a way to explore the electroweak form factors for both the proton and neutron; that is, equivalently, both isoscalar and isovector form factors of the nucleon may be explored in this way. We have introduced specific asymmetries involving protons with neutrons where the sensitivities to the hard-to-determine form factors are particularly large. Exploiting such asymmetries also has the advantage of minimizing systematic errors in future measurements. We have found that the cross sections are very sensitive to the isoscalar axial-vector form factor as characterized by its strength parameter $g_A^s$, and sensitive but less so to the magnetic strangeness form factor characterized by its strength parameter $\mu_s$. In the kinematic region explored in this study little sensitivity was found to the electric strangeness form factor with its strength parameter $\rho_s$.

The present study has been focused on neutral-current scattering of neutrinos, although some attention was paid to the differences that occur when anti-neutrinos are employed. In particular, the vector-vector plus axial-axial contributions remain unchanged, while the vector-axial contributions change sign in going from neutrinos to anti-neutrinos. Again an asymmetry can be created by forming the ratio of the semi-inclusive cross sections ({\it i.e.,} neutrino minus anti-neutrino over their sum). From an exploratory study done involving both neutrinos and anti-neutrinos we see rather similar results in the two cases, although with somewhat different sensitivities to the underlying nucleon form factors. Clearly having both types of beam could yield more information about the form factors while potentially minimizing some experimental issues of a systematic nature.

Finally, we note that the present study is centered on the underlying theory involved, both the two-body nuclear theory and the hadronic physics that could be explored in future experiments. This work can serve as a basis for developing those future experiments, perhaps using heavy water or some other target that is rich in deuterium, although that is a job for those who are interested in pursuing such experimental issues and is not part of our focus. Upon moving towards a conceptual experiment one would hope that a deuterium-rich target/detector could be realized where both charge-changing and neutral-current disintegration reactions could be measured in concert.

{\bf Acknowledgments}: This work was
supported in part by funds provided by the National Science Foundation under
grants No. PHY-1614460 and PHY-1913261 (S. J.), by Jefferson Science Associates, LLC under U.S. DOE Contract DE-AC05-06OR23177 and by U.S. DOE Grant DE-FG02-97ER41028 (J. W. V. O.),  and in part by the Office of Nuclear Physics of the US Department of Energy under Grant Contract DE-FG02-94ER40818 (T. W. D.).

\bibliography{NC}

\end{document}